\documentclass{article}

\usepackage[english]{babel}

\usepackage[a4paper,top=2cm,bottom=2cm,left=3cm,right=3cm,marginparwidth=1.75cm]{geometry}

\usepackage{amsmath}
\usepackage{graphicx}
\usepackage[colorlinks=true, allcolors=blue]{hyperref}
\usepackage{comment}
\usepackage{placeins}
\usepackage{hyperref}
\usepackage[numbers,sort&compress]{natbib}

\newcommand{\GeV}{\ensuremath{\text{Ge\kern -0.1em V}}}
\newcommand{\MeV}{\ensuremath{\text{Me\kern -0.1em V}}}
\newcommand{\keV}{\ensuremath{\text{ke\kern -0.1em V}}}

\title{ParaFlow: fast calorimeter simulations \\ parameterized in upstream material configurations}
\author{Johannes Erdmann, Jonas Kann, Florian Mausolf, Peter Wissmann}

\date{\small
  RWTH Aachen University, III. Physikalisches Institut A, Aachen, Germany
  }

\begin{document}
\maketitle

\begin{abstract}
We study whether machine-learning models for fast calorimeter simulations can learn meaningful representations of calorimeter signatures that account for variations in the full particle detector's configuration.
This may open new opportunities in high-energy physics measurements, for example in the assessment of systematic uncertainties that are related to the detector geometry, in the inference of properties of the detector configuration, or in the automated design of experiments.
As a concrete example, we parameterize normalizing-flow-based simulations in configurations of the material upstream of a toy calorimeter.
We call this model ParaFlow, which is trained to interpolate between different material budgets and positions, as simulated with \textsc{Geant4}.
We study ParaFlow's performance in terms of photon shower shapes that are directly influenced by the properties of the upstream material, in which photons can convert to an electron-positron pair.
In general, we find that ParaFlow is able to reproduce the dependence of the shower shapes on the material properties at the few-percent level with larger differences only in the tails of the distributions.
\end{abstract}

\section{Introduction}

The simulation of the detector response with \textsc{Geant4}~\cite{GEANT4:2002zbu} is an essential ingredient in almost all measurements in current high-energy physics, such as at the Large Hadron Collider (LHC).
However, the computational cost that is associated with the production of simulated samples is large~\cite{CERN-LHCC-2020-015} and driven by the simulation of the shower development in the calorimeters.
Alternatives that require less resources (``fast simulations'') are hence needed.
Traditionally, fast simulations use parameterizations of the calorimeter response~\cite{ATLAS:1300517,ATL-SOFT-PUB-2014-001,ATLAS:2021pzo,Abdullin:2011zz,Hildreth:2297284}.
Generative models based on deep learning, however, have been proposed as a promising approach for fast calorimeter simulations~\cite{Paganini:2017dwg}.
They have gained much attention (for recent reviews see Refs.~\cite{Hashemi:2023rgo,Ahmad:2024dql}) and first models are now in use at LHC experiments~\cite{ATLAS:2021pzo,Barbetti:2023bvi}.

These generative models are surrogates that are trained to reproduce predictions based on the ``full simulation'' with \textsc{Geant4} as closely as possible.
Various generative machine learning approaches have been shown to be capable of simulating calorimeter showers of high quality, including models based on
variational auto-encoders~\cite{ATLAS:2022jhk,deja2020endtoendsinkhornautoencodernoise,Buhmann:2020pmy,Buhmann:2021lxj,Buhmann:2021caf,Diefenbacher:2023prl,Hariri:2021clz,AbhishekAbhishek:2022wby,Cresswell:2022tof,Liu:2024kvv},
generative adversarial networks~\cite{deOliveira:2017pjk,Paganini:2017dwg,deOliveira:2017rwa,Paganini:2017hrr,Khattak:2021ndw,8451587,Vallecorsa:2019ked,Belayneh:2019vyx,Musella:2018rdi,Chekalina:2018hxi,Diefenbacher:2020rna,Jaruskova:2023cke,FaucciGiannelli:2023fow,Erdmann:2018jxd,Carminati:2018khv,Erdmann:2023ngr},
flow-based models~\cite{Krause:2021ilc,Krause:2021wez,Krause:2022jna,Pang:2023wfx,CaloPointFlowI,Schnake:2024mip,Buss:2024orz,Dreyer:2024bhs,Ernst:2023qvn,Favaro:2024rle},
diffusion models~\cite{Mikuni:2022xry,Mikuni:2023tqg,Buhmann:2023bwk,Buhmann:2023kdg,Diefenbacher:2023flw,Amram:2023onf,Acosta:2023zik,Kobylianskii:2024ijw,Kobylianskii:2024sup,madula_mikuni},
and autoregressive models~\cite{Lu:2020npg,Liu:2022dem,Liu:2023lnn,Diefenbacher:2023vsw,Buckley:2023daw},
following the taxonomy of Ref.~\cite{Hashemi:2023rgo}.
A detailed comparison of the performance in terms of generation quality and timing benchmarks for a large range of models can be found in the HEP community challenge on calorimeter simulations~\cite{Krause:2024avx}.

Most of these works include models that have been trained to reproduce signatures in a specific detector.
So far, only a few studies have gone beyond.
In Refs.~\cite{Smith:2024lxz,Liu:2023lnn,Liu:2022dem}, a ``geometry-aware'' approach was proposed, which allows for varying sizes of the calorimeter readout cells.
Alternatively, in Refs.~\cite{Buhmann:2023bwk,Buhmann:2023kdg,CaloPointFlowI,Schnake:2024mip}, ``geometry-independent'' models were developed that generate point clouds in a given detector material and are hence independent of the exact readout geometry.
In Ref.~\cite{Raikwar:2024peb}, a ``detector-agnostic'' model was studied that is intended to be easily adaptable to other calorimeters.
This was inspired by Ref.~\cite{Salamani:2023ttx}, which explored meta-learning to interpolate between calorimeters with different materials.
In addition, an approach called ``geometry adaption'' was pursued in Ref.~\cite{Amram:2023onf}, which relies on a latent mapping of the calorimeter geometry.
All of these works have focussed on relaxing the assumptions on the calorimeter geometry.
While some of these works addressed aspects related to the readout segmentation for a given distribution of detector material~\cite{Smith:2024lxz,Liu:2023lnn,Liu:2022dem,Buhmann:2023bwk,Buhmann:2023kdg}, others aimed to learn the translation between different distributions of material in the calorimeters for a small set of scenarios~\cite{Raikwar:2024peb,Salamani:2023ttx}.
Although calorimeter signatures are directly influenced by the amount of material (``material budget'') in front of the calorimeter (``upstream material''), the dependence of fast calorimeter simulations on other parts of the full particle detector has not gained much attention.

We ask whether generative models for calorimeter simulations can learn a meaningful representation of variable configurations for the full particle detector, i.e., in a single generative model.
Such a model would have important implications for high-energy physics experiments: it could allow for a more efficient and more fine-grained evaluation of systematic uncertainties that are related to the detector configuration and which can be important~\cite{ATLAS:2023owm,CMS:2020xrn}; it could be used to estimate parameters of the detector configuration directly from data; and it could serve the automated design of particle physics experiments~\cite{MODE:2021yid,MODE:2022znx,Shirobokov:2020tjt,Neubuser:2021uui,Aehle:2023wwi,Aehle:2024ezu,Strong:2023oew,MODE:2025zir,Wozniak:2025ttb}, where surrogate models can be used for gradient-based end-to-end optimizations.

We study this for a CMS-like toy calorimeter with an iron block with variable thickness and position in front, which serves as a proxy for the upstream material, i.e., the tracking detector, support material etc.
The basic concept of the study is independent of the chosen generative model.
As a state-of-the-art example, we use an architecture that is closely inspired by the CaloFlow model~\cite{Krause:2021ilc}, which is based on normalizing flows (NFs)~\cite{Tabak, papamakarios2021normalizing}.
We study photon signatures, as they are directly affected by the amount and position of the upstream material.
The more upstream material is present, the more conversions to electron-positron pairs are expected to occur, resulting in wider showers in the calorimeter on average.
In addition, the further away the material is positioned from the calorimeter, the longer the flight distance of electrons and positrons in the magnetic field, which again results in wider showers.

We choose a simple two-parameter model for the upstream material with one parameter for the thickness of the iron block and another parameter for its distance to the calorimeter.
We train a single NF, which we call ParaFlow, on photons simulated with \textsc{Geant4} to learn the dependence of the photon's calorimeter signature as a function of these parameters.
We then compare ParaFlow's predictions to the expectation from \textsc{Geant4} simulations to assess the quality of the surrogate model.
We finally conclude whether ParaFlow is able to learn a meaningful representation of the calorimeter signature as a function of the detector configuration parameters.

\section{Simulated samples}
\begin{figure}[t]
    \centering
    \includegraphics[width=0.54\linewidth]{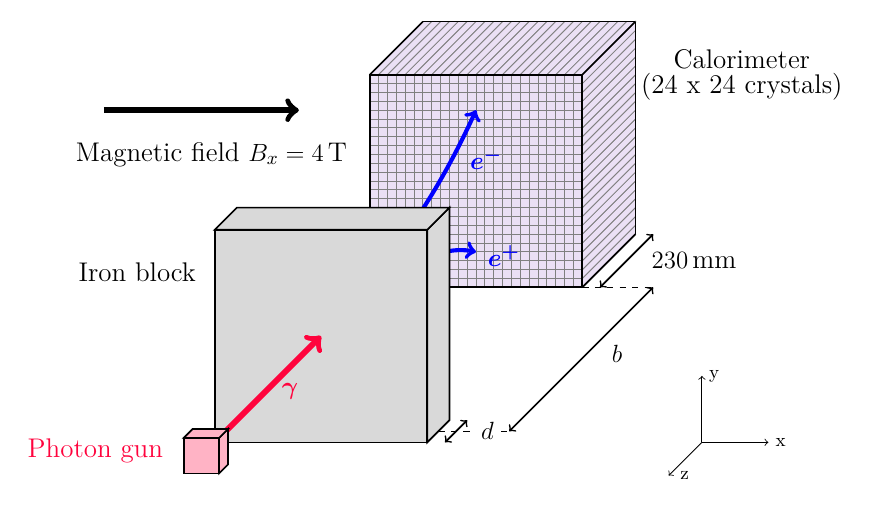}
    \caption{Sketch of the toy detector. The particle gun shoots photons perpendicular to the PbWO$_4$ calorimeter. The photons may undergo conversion in the iron block, which represents the material upstream of the calorimeter, producing an $e^+e^-$ pair. The whole volume contains a homogenous magnetic field.
    The distance $b$ between the centre of the iron block and the calorimeter front face, as well as the thickness $d$ of the iron block are variable.
    }
    \label{fig:simulation_setup}
\end{figure}

We simulate a toy calorimeter that is inspired by the electromagnetic barrel calorimeter of the CMS detector~\cite{CMS:2008xjf}. 
The simulation is based on the framework of Ref.~\cite{Paganini:2017dwg} with \textsc{Geant4}~\texttt{11.2.2} and builds on the CMS-like toy calorimeter from Ref.~\cite{Erdmann:2023ngr}. 
We simulate PbWO$_4$ scintillating crystals with a length of $230\,\text{mm}$ and a front face of $22\times22\,\text{mm}^2$. 
The front of the calorimeter is placed at a distance of $1.29\,\text{m}$ from a \textsc{Geant4} particle gun that simulates photons. 
These photons have a uniformly distributed energy between $20\,\GeV$ and $100\,\GeV$, where the direction of the photons is perpendicular to the front face of the detector. 
This energy range is chosen to represent typical photon energies analysed at the LHC, for example, in measurements targetting $H\to\gamma\gamma$ decays.
To distribute the energy deposition in the detector such that not all photons hit the toy calorimeter centrally, the position of the source is smeared. 
For that, a Gaussian distribution with a width of $44\,\text{mm}$ is chosen.
This corresponds to the width of two crystals. 
A magnetic field with a field strength of $B=4\,\text{T}$ is added in $x$-direction, as shown in Fig.~\ref{fig:simulation_setup}. 
The magnetic field widens the photon showers in one direction, as in typical particle detectors, and its strength is chosen to be close to the field strength of the CMS solenoid.
To mimic the effect of material upstream of the calorimeter on the photon conversion rate and on the shower formation, an iron plate with a front surface of $1\times1\,\text{m}^2$ is added between the source and the calorimeter.
Its distance to the calorimeter, $b$, and its thickness, $d$ are parameters of the simulation setup.
These parameters were initialised separately for each photon, i.e., a separate run was started with different parameter settings per photon.
The parameters were varied uniformly in the ranges $50\,\mathrm{cm} \leq b \leq 90\,\mathrm{cm}$ and $0.5 \,X_0 \leq d \leq 1.5 \,X_0$, where $X_0 = 1.757\,\text{cm}$ is the radiation length of iron.
The central values of the intervals are inspired by the CMS detector~\cite{CMS:2014pgm}.
The interval ranges are chosen such that the variations have a significant impact on the shower properties within the statistics of the simulated sample.
We did not simulate noise in the calorimeter, nor other aspects of digitization.
Approximately $1,000,000$ photon showers were simulated in total, of which $60\%$ were used for training, $10\%$ for validation, and $30\%$ for testing purposes.

\section{Machine-learning architecture}

Normalizing flows map the input data to a base space via a learnable bijective transformation.
Generated samples in the input space are obtained by sampling from the known distribution in the base space, typically a multidimensional Gaussian distribution.
As in Ref.~\cite{Krause:2021ilc}, we use masked auto-regressive flows (MAFs)~\cite{papamakarios2017masked} with monotonic rational quadratic spline (RQS) transformations~\cite{durkan2019neural}.
The parameters of the RQS transformations are calculated by MADE blocks~\cite{germain2015made}.
In Ref.~\cite{Krause:2021ilc}, two NFs were used, where the first predicts the layer-wise energy deposition in a multi-layer calorimeter, and the second generates the distribution of the energy within a layer.
As our calorimeter consists of a single layer, we use a single NF.

\begin{figure}[t]
    \centering
    \includegraphics[width=\linewidth]{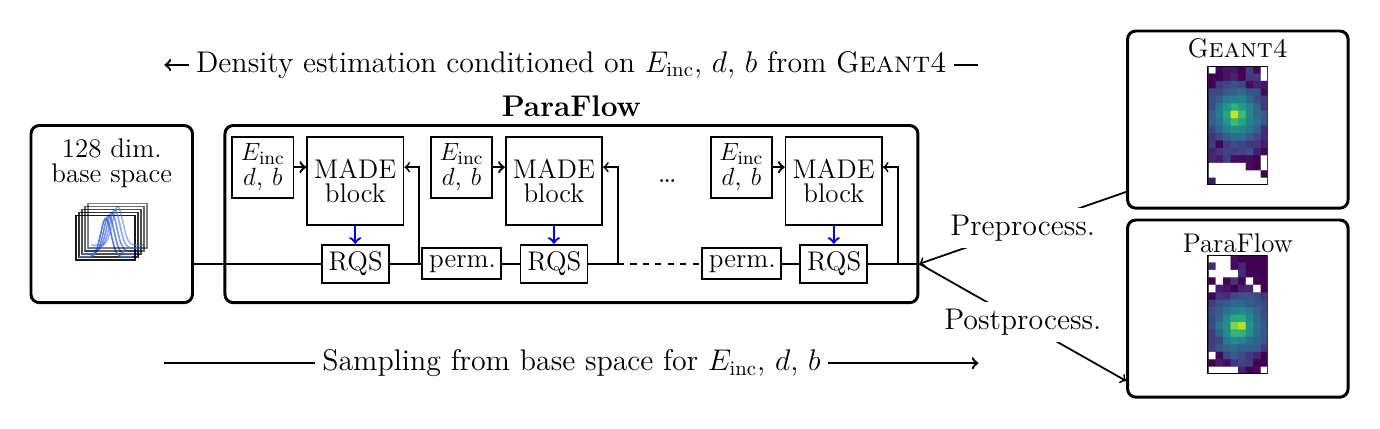}
    \caption{Illustration of the NF architecture. The figure design is inspired by Ref.~\cite{Krause:2021ilc}.}
    \label{fig:architecture}
\end{figure}

The architecture is illustrated in Fig.~\ref{fig:architecture}.
The energy depositions of \textsc{Geant4}-simulated photons in the calorimeter crystals (``calorimeter images'') are mapped to a 128-dimensional base space, conditioned on the incoming photon energy $E_\text{inc}$ and the material variables $d$ and $b$.
The calorimeter images are flattened before being passed to the NF, appended by the conditions. Between the MADE blocks, the variables are randomly permuted.

The main hyperparameters are the number of MADE blocks, the structure of the multi-layer perceptrons (MLPs) within the MADE blocks, and the number of bins for the RQS transformations.
We build the NF with ten MADE blocks and ten spline bins.
These parameters were optimised in a coarse grid search based on the area under the ROC curve of a ResNet~\cite{he2015deepresiduallearningimage} classifier that was trained to discriminate generated from \textsc{Geant4}-simulated photons.
The MADE blocks consist of MLPs with two hidden layers and 196 nodes each, activated with ReLU.
Wider MLPs were not found to improve the performance.
The resulting model has approximately 8~million parameters.

The calorimeter images are preprocessed in the following way:
The $24\times24$ simulated energy depositions in the calorimeter crystals are cropped to the $8\times 16$ window that contains the maximum energy sum.
The window's asymmetry accounts for the impact of the magnetic field on the shower formation.
Even for showers generated with the thickest iron plate and largest distance from the calorimeter, i.e., where the widest showers are expected, on average less than $1\%$ of the energy is deposited outside of the window.
As in Ref.~\cite{Krause:2021ilc}, we add a small noise, which we choose to be uniformly distributed between $0$ and $20\,\keV$.
The energy deposits in the individual crystals are then divided by the incoming photon energy $E_{\mathrm{inc}}$ and processed by a logit transformation, as in Ref.~\cite{Krause:2021ilc}.
The conditions $c_i$ are logarithmically transformed, such that they are mapped to the domain $[-1,1]$:
\begin{equation}
    c_i \quad \rightarrow \quad 2\cdot \frac{\mathrm{log}(c_i - (c_{i, \text{min}}-1))}{\mathrm{log}(c_{i, \text{max}} - (c_{i, \text{min}}-1))} - 1 \quad \in [-1,1], \; \text{for} \; c_i \in [c_{i, \text{min}}, c_{i, \text{max}}]
    \label{eq:condition_scaling}
\end{equation}

The NFs are implemented in PyTorch~\cite{paszke2019pytorch} using the Zuko package~\cite{rozet2022zuko}.
The training is performed with a batch-size of $1000$ using the Adam optimizer~\cite{Kingma:2014vow} with an initial learning rate of $10^{-3}$ and a cosine annealing learning-rate scheduler with warm restarts \cite{CosineAnnealing}. The restarting learning rate is reduced by a factor of $0.5$ on plateaus of the validation loss with a patience of $10$ epochs.
The training is stopped if the validation loss did not improve over $15$~epochs.

\section{Results}

\subsection{Qualitative comparisons}

\begin{figure}[h]
    \centering
    \includegraphics[width=0.8\linewidth]{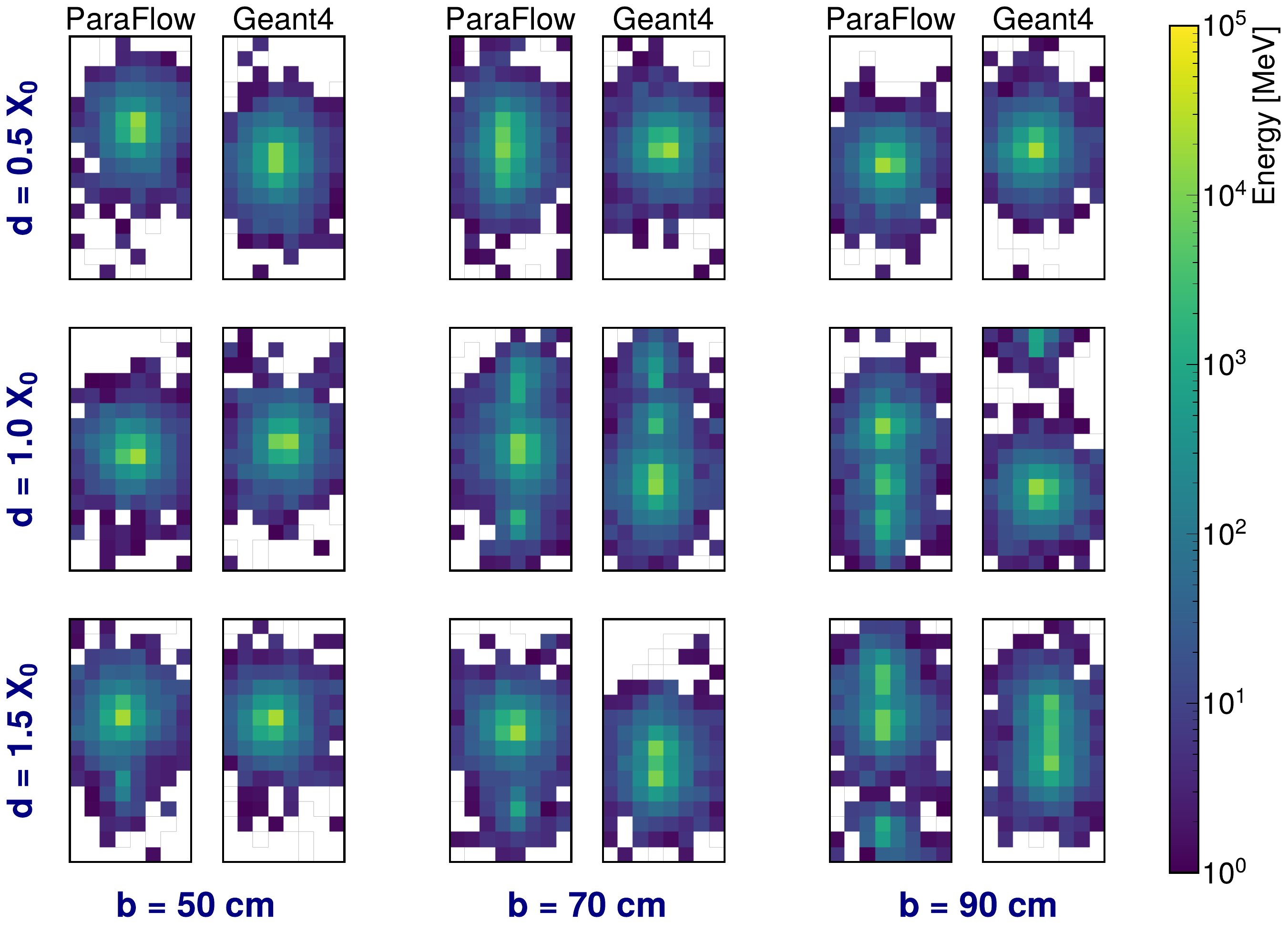}
    \caption{Generated calorimeter images of $30\,\GeV$ photons for varying material conditions $d$ and~$b$, compared with randomly selected \textsc{Geant4} images with very similar values of $d$ and $b$. The random selection implies that images are not expected to closely match between \textsc{Geant4} and generated images.}
    \label{fig:shower_samples}
\end{figure}

Figure~\ref{fig:shower_samples} shows example generated calorimeter images for varying values of material budget ($d$) and position ($b$) upstream of the calorimeter, for an energy of $E_{\mathrm{inc}} = 30\,\GeV$.
These are compared with randomly selected \textsc{Geant4} images with very similar values of $d$ and $b$.
As expected, the generated and the \textsc{Geant4}-simulated photons show a trend towards more energy clusters in the image when the material budget increases, as the conversion probability rises.
The clusters also tend to be separated more strongly when the distance of the iron block and the calorimeter increases, because of the longer distance that electrons and positrons from such conversions travel in the magnetic field.

\begin{figure}[h]
    \centering
    \includegraphics[width=0.8\linewidth]{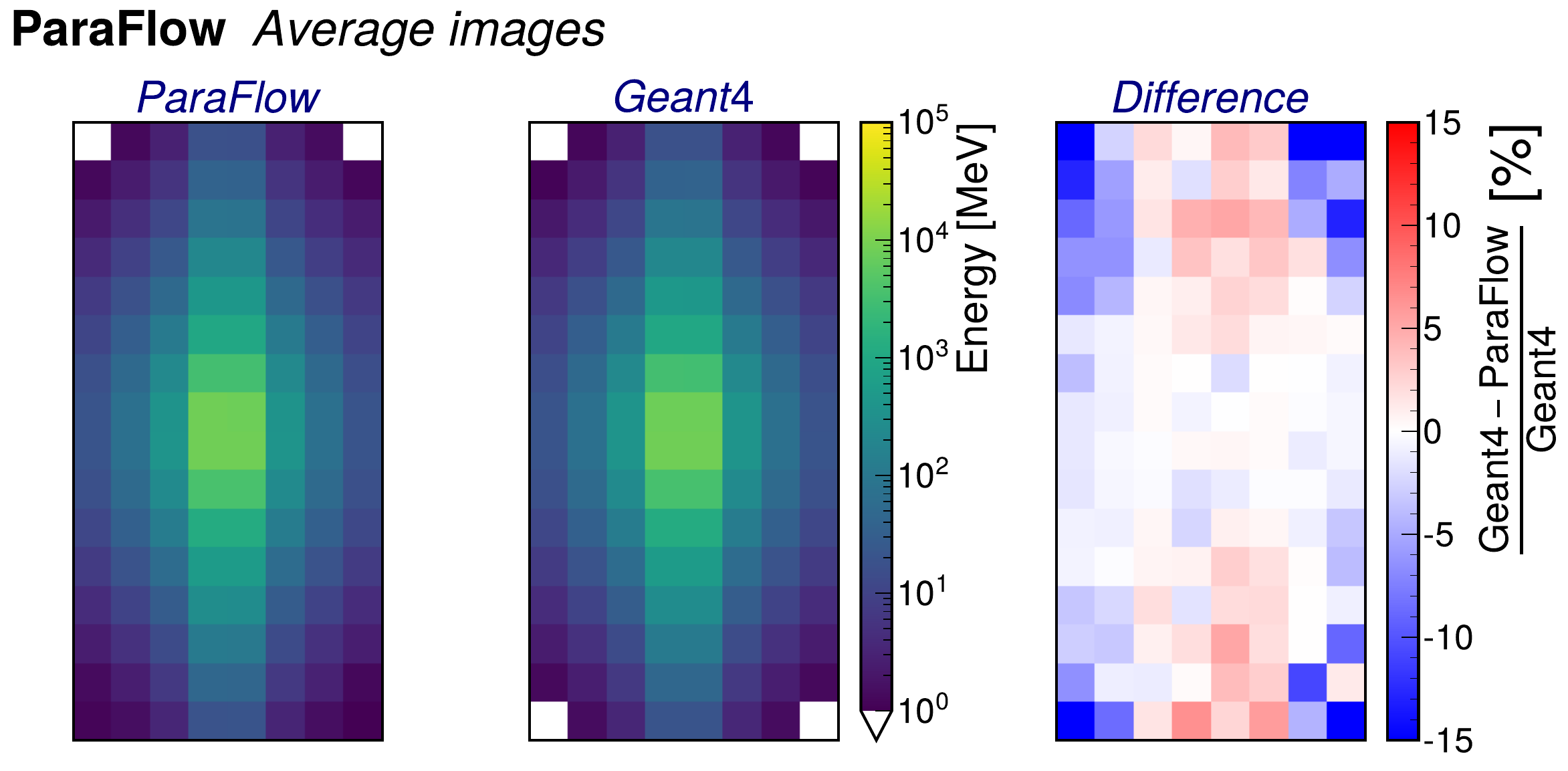}
    \caption{Averages of \textsc{Geant4}-simulated and ParaFlow-generated calorimeter images, taken across all energies and detector conditions. On the right, their relative difference is shown.}
    \label{fig:shower_averages}
\end{figure}

Figure~\ref{fig:shower_averages} shows the average of all \textsc{Geant4} calorimeter images in the test dataset ($300,000$ photons), the average of $1,000,000$ calorimeter images generated with ParaFlow with uniform conditions, as well as their relative difference.
Among the central $4 \times 4$ crystals, the highest relative deviation amounts to $2\%$.
For the crystals with larger distances from the centre, the deviations mostly remain below $15\%$.
Since the average energies in these crystals is of the order of $10\,\MeV$, they are expected to only contribute very little to reconstructed shower properties, such as the shower shape observables discussed below.

\subsection{Distributions of shower shape observables}

We compare the predictions obtained from the ParaFlow samples with the corresponding~\textsc{Geant4} results using shower shape observables.
We quantify the agreement using the separation power of histograms, also used in Ref.~\cite{Krause:2024avx}, and given by 
\begin{equation}
    S(h_1, h_2) = \frac{1}{2} \sum_i \frac{(h_{1,i} - h_{2,i})^2}{h_{1_i} + h_{2,i}} \, ,
\end{equation}
where $h_{n,i}$ is the count of the $i^{\mathrm{th}}$ bin of the $n^{\mathrm{th}}$ ($n = 1, 2$) normalised histogram. A separation power of $S = 0$ indicates perfectly matching histograms, and $S = 1$ quantifies vanishing agreement.

\begin{figure}[t]
    \centering
    \begin{minipage} {0.45 \textwidth}
        \centering
        \includegraphics[width = \textwidth]{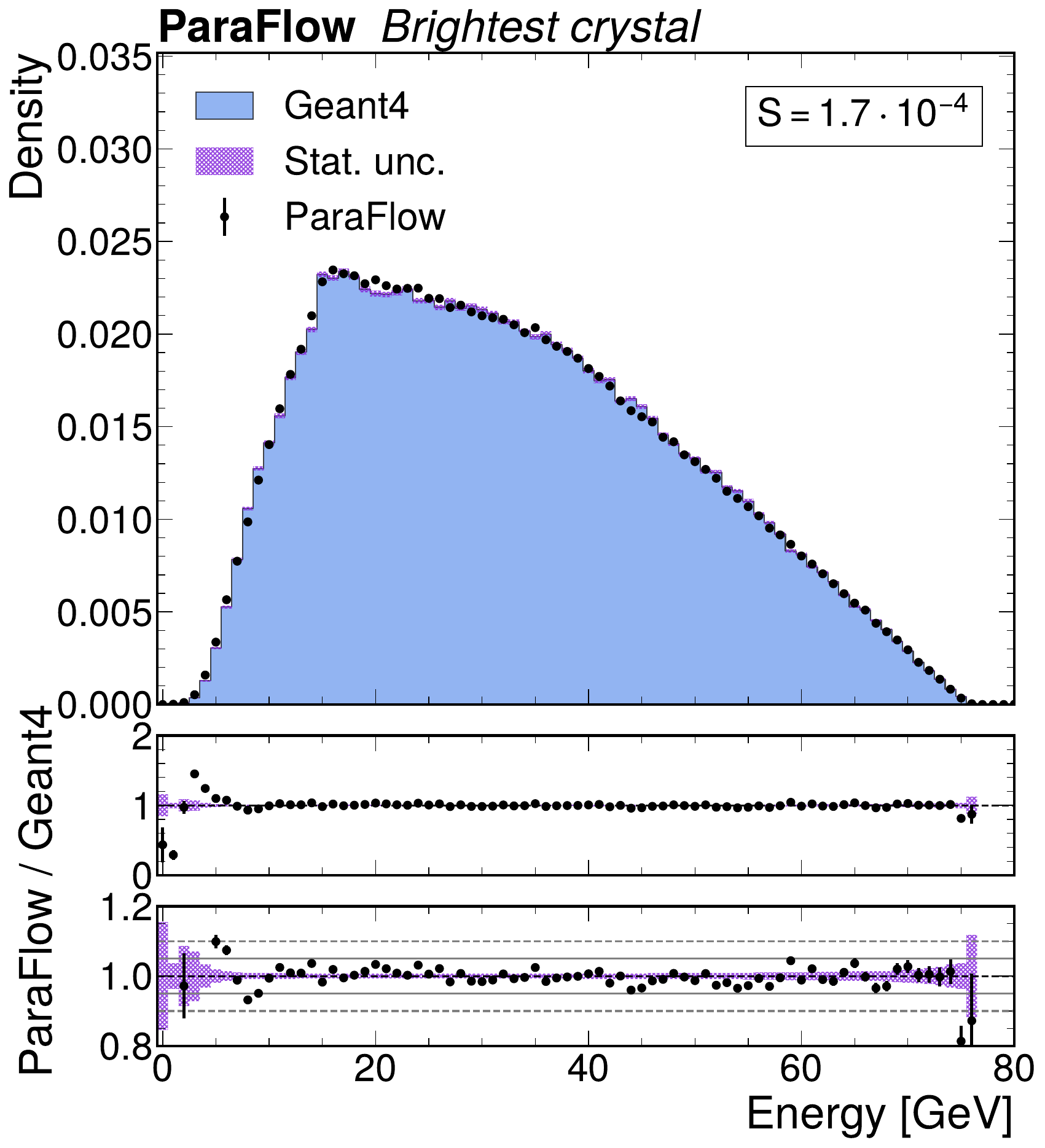}
    \end{minipage}
    \begin{minipage} {0.45 \textwidth}
        \centering
        \includegraphics[width = \textwidth]{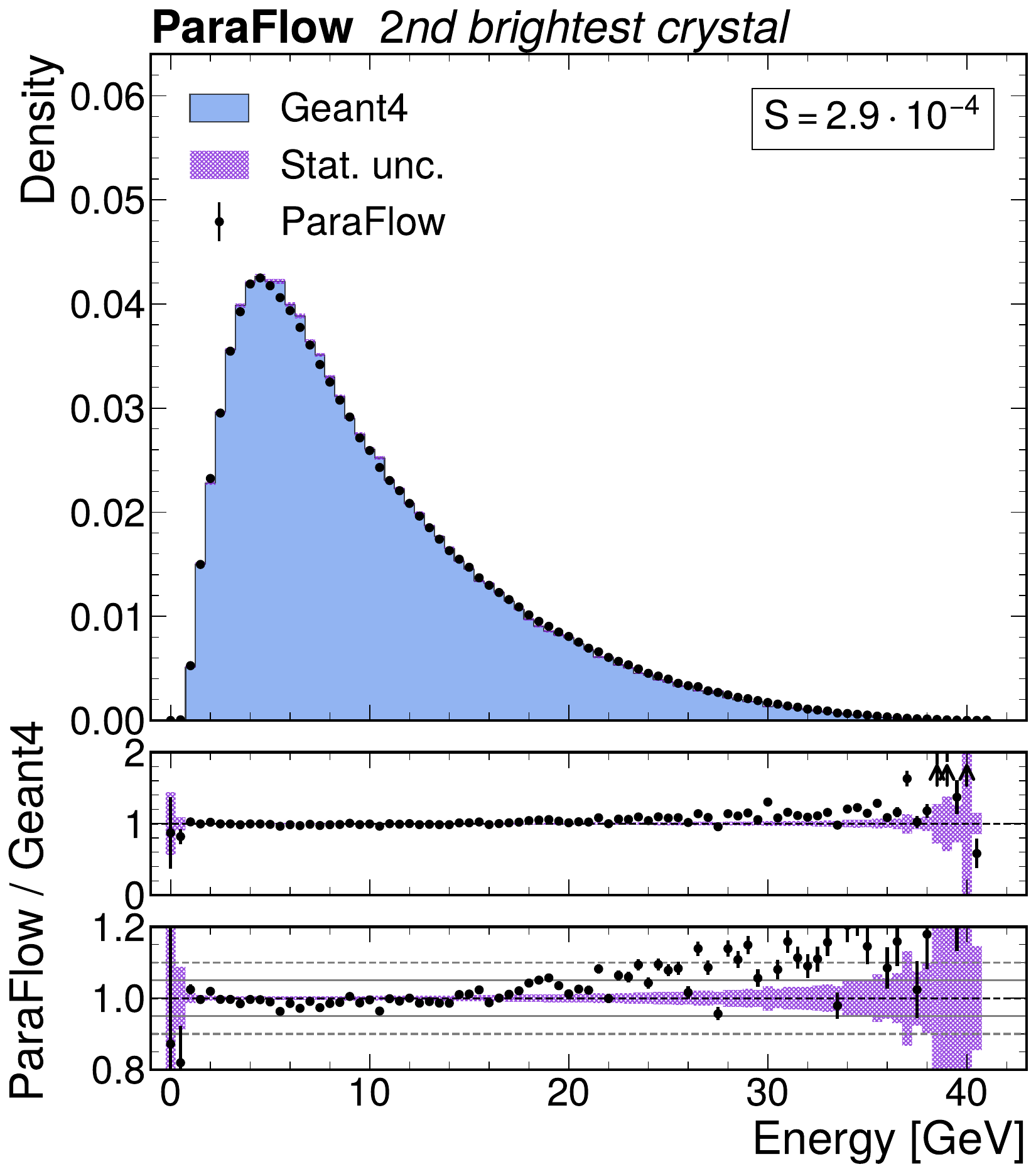}
    \end{minipage}
    \caption{Distributions of the energy in the brightest (left) and second-brightest (right) crystal in the calorimeter in the full dataset, i.e., integrated over the two parameters that describe the material upstream of the calorimeter, as well as over photon energies. The plots below show the ratio of ParaFlow over \textsc{Geant4} on two different $y$-axis scales.
    Both observables are largely insensitive to the variations of the upstream material and are used to assess the overall performance of ParaFlow.}
    \label{fig:insensitive_histograms}
\end{figure}

To quantify the overall performance of ParaFlow in describing the \textsc{Geant4} samples, the energy distributions in the brightest and second brightest crystals are shown in Fig.~\ref{fig:insensitive_histograms}.
These observables are largely insensitive to variations in the material upstream of the calorimeter and allow us to assess whether our model shows the good description that is expected for a CaloFlow-based model.
In contrast to Ref.~\cite{Krause:2021ilc}, however, we did not normalise the energies in Fig.~\ref{fig:insensitive_histograms} by the total deposited energy in our one-layer calorimeter,
as we expect ParaFlow to also learn the overall energy.
For the energy distribution of the brightest crystal, we see a good agreement between the ParaFlow samples and the \textsc{Geant4} data. The deviations remain below $5\%$ over almost the entire range. Similarly, we find deviations below $5\%$ in the bulk of the distribution of the second-brightest crystal. Deviations occur in the high-energy tail that may be due to an insufficient representation of shower images with a high-energy second-brightest crystal in the training data. Overall, the histograms show a good separation power of $\mathcal{O} (10^{-4})$, which indeed indicates good performance of our model.

To assess whether ParaFlow is able to accurately capture the dependence of the detector simulation on the material upstream of the calorimeter, we divide the ParaFlow test set and the \textsc{Geant4}-generated dataset into equal-sized subsets, either divided by $d$ or $b$ (i.e., inclusive in the other variable).
As in Fig.~\ref{fig:shower_averages}, the ParaFlow test set is comprised of 1,000,000 ParaFlow-generated images with uniform conditions, and the \textsc{Geant4} test set contains 300,000 simulated photons.
The subsets are defined by the boundaries $[0.5, 0.75, 1.0, 1.25, 1.5] \, X_0$ for $d$, and $[50, 60, 70, 80, 90] \, \mathrm{cm}$ for~$b$.
We use the following shower shape observables to assess ParaFlow's performance in describing the dependence on the material.
These observables are particularly sensitive to the location and probability of conversions in the material upstream of the calorimeter:
\begin{itemize}
    \item The width of the shower is defined as
        \begin{equation}
            w = \frac{\sum_i r_i E_i}{\sum_i E_i}\, ,
        \end{equation}
    with $E_i$ the energy deposited in the $i^{\mathrm{th}}$ crystal and $r_i$ its Euclidean distance to the barycentre of the shower in units of crystal widths.
    \item The variable $R_9$ is defined as the sum of the energies in the $3 \times 3$ crystals surrounding the crystal with the highest energy deposition, divided by the total energy, $E_{\mathrm{tot}} = \sum_i E_i$:
        \begin{equation}
            R_9 = \frac{E_{3\times3}}{E_{\mathrm{tot}}}
        \end{equation}
    This quantity is commonly used in the CMS experiment to distinguish converted from unconverted photons~\cite{CMS:2015myp, CMS:2020uim}.
    Lower values are expected for converted photons, in particular for early conversions, i.e., large values of $b$.
    \item We use the DBSCAN clustering algorithm~\cite{DBSCAN} to define the number of energy clusters in the calorimeter image.
    This algorithm relies on three parameters: $\varepsilon$, the maximum distance between points to be considered neighbours, \texttt{minPts}, the minimum number of points needed to form a cluster, and the energy threshold $E_{\mathrm{min}}$ to filter out low-energy crystals. Crystals with $E > E_{\mathrm{min}}$ and which meet the \texttt{minPts} requirement within the radius $\varepsilon$ form clusters. We choose $\varepsilon = 1.5$ crystal widths to include only direct neighbours, \texttt{minPts}$ = 1$, and $E_{\mathrm{min}} = 400\,\MeV$, to stay above typical experimental noise levels.
\end{itemize}

\begin{figure}[p]
        \centering
        \begin{minipage} {0.45 \textwidth}
            \centering
            \includegraphics[width = \textwidth]{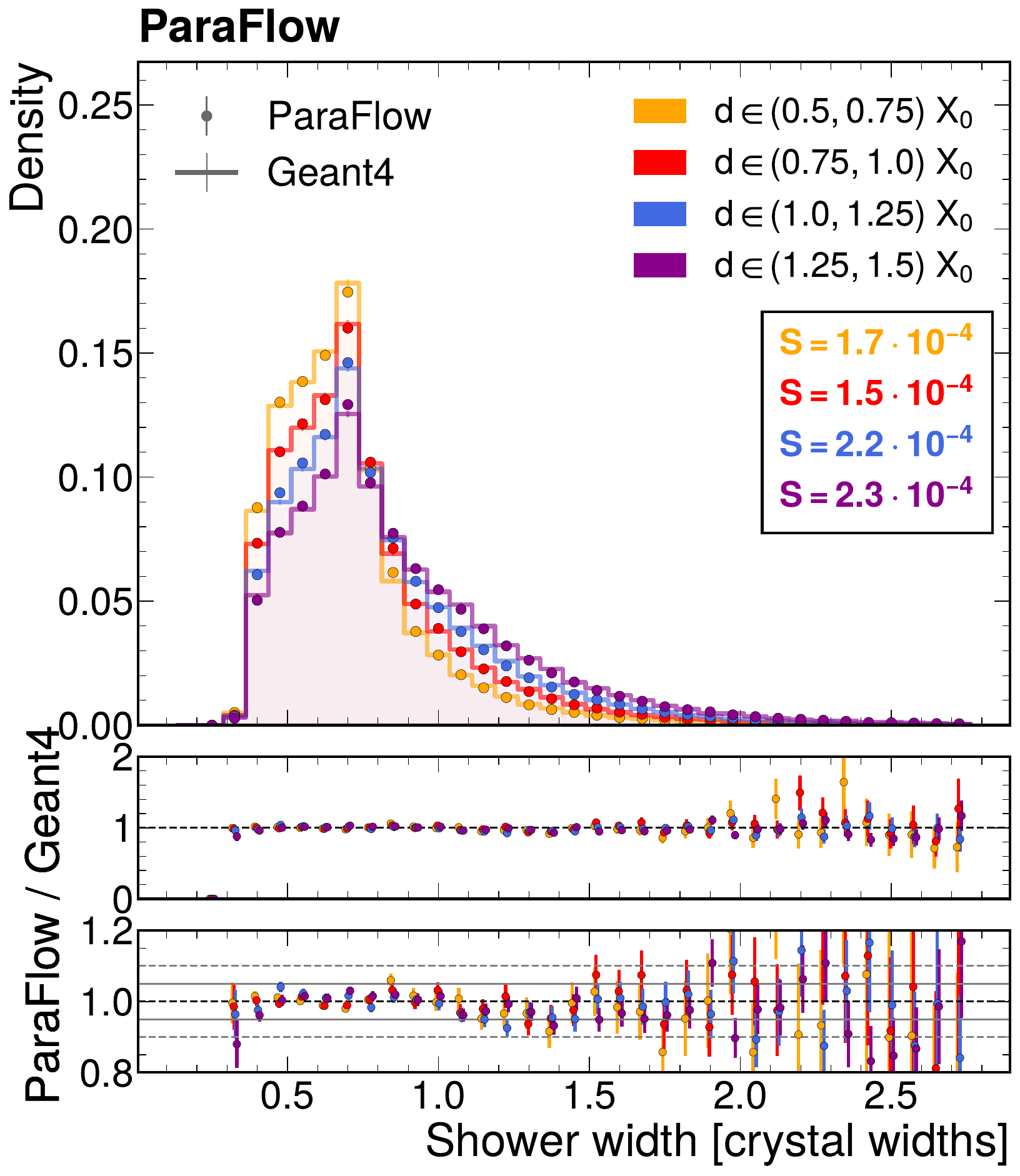}
        \end{minipage}
        \begin{minipage} {0.45 \textwidth}
            \centering
            \includegraphics[width = \textwidth]{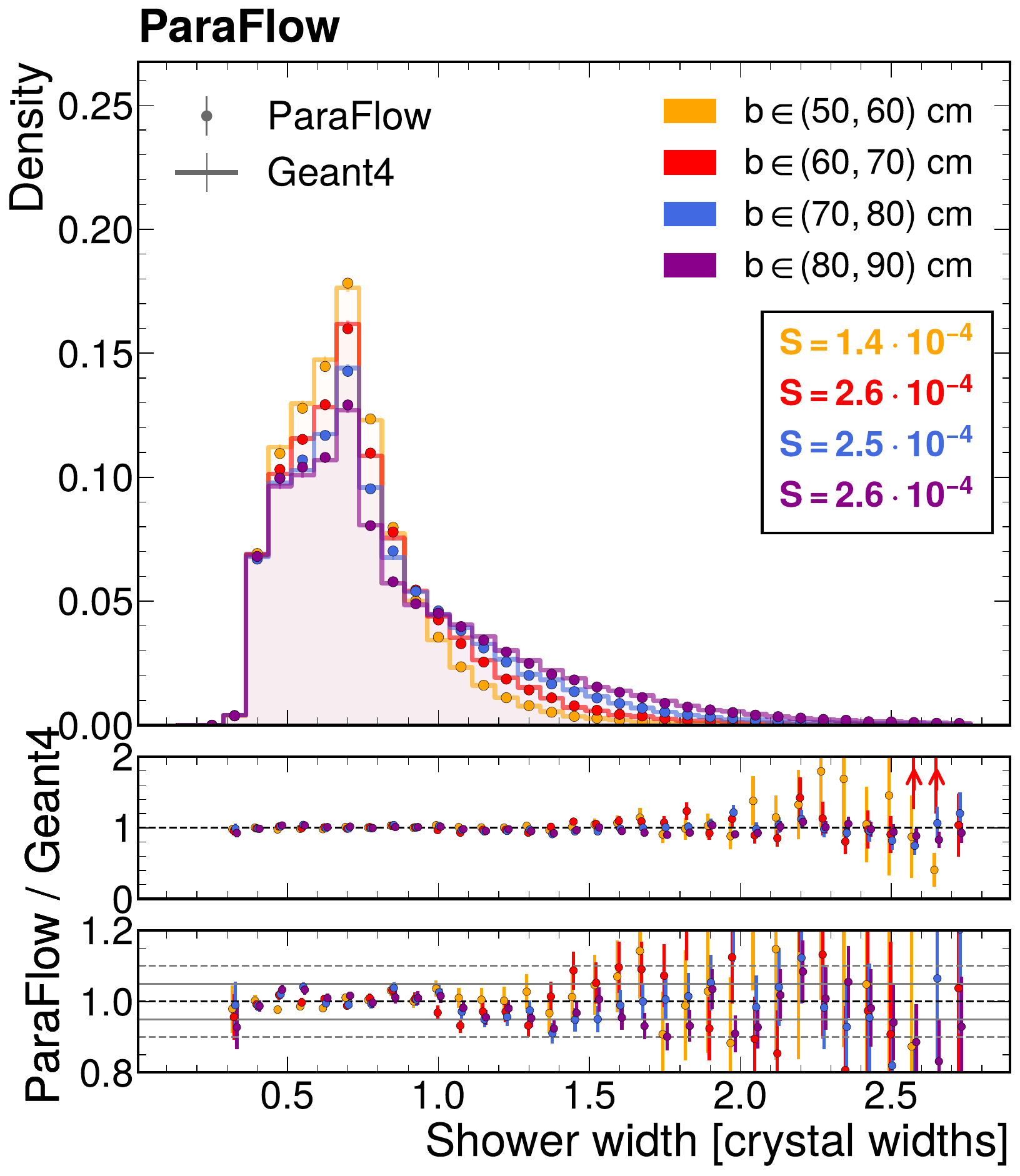}
        \end{minipage}
        \caption{Distributions of the shower width, $w$, subdivided according to the values of the material parameters: distributions for varying material budget $d$ (left) and for varying material distance $b$ (right).
        The plots below show the ratio of ParaFlow over \textsc{Geant4} on two different $y$-axis scales.
        }
        \label{fig:histograms_shower_width}
        \vspace{1cm}
        \begin{minipage} {0.45 \textwidth}
            \centering
            \includegraphics[width = \textwidth]{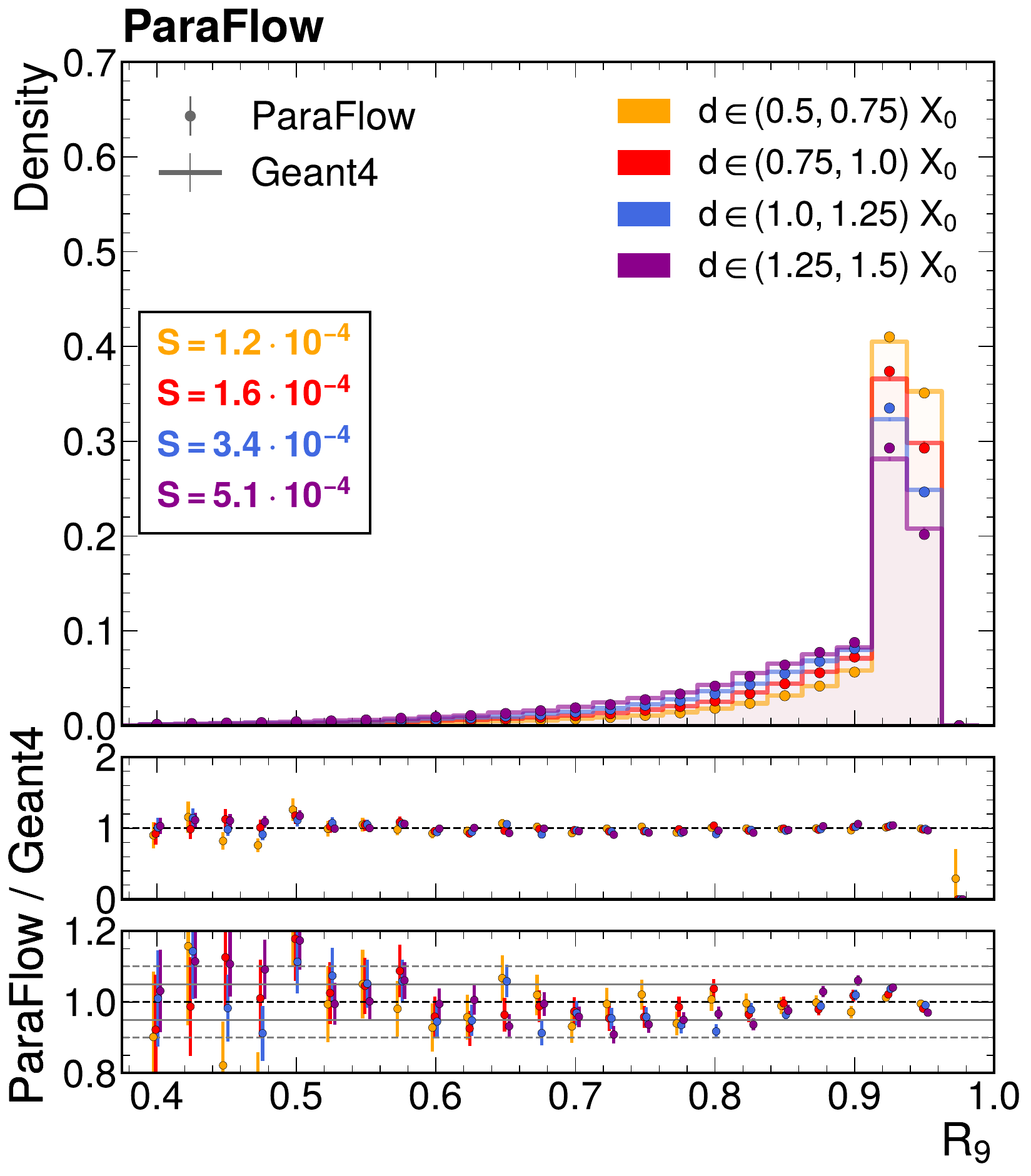}
        \end{minipage}
        \begin{minipage} {0.45 \textwidth}
            \centering
            \includegraphics[width = \textwidth]{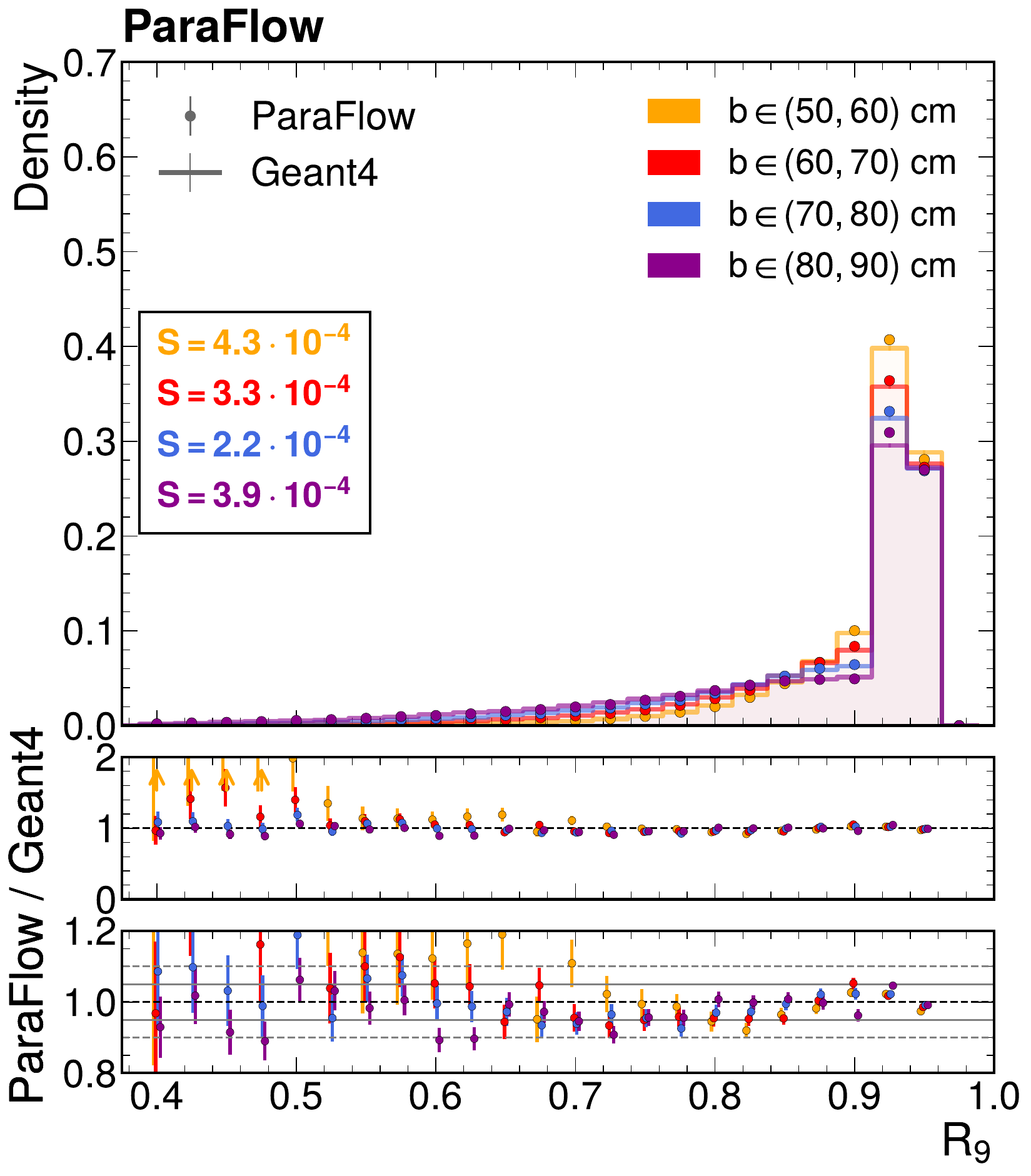}
        \end{minipage}
        \caption{Distributions of $R_9$, subdivided according to the values of the material parameters: distributions for varying material budget $d$ (left) and for varying material distance $b$ (right).
        The plots below show the ratio of ParaFlow over \textsc{Geant4} on two different $y$-axis scales.
        }
        \label{fig:histograms_r9}
    \end{figure}

\begin{figure}[h!]
        \centering
        \begin{minipage} {0.45 \textwidth}
            \centering
            \includegraphics[width = \textwidth]{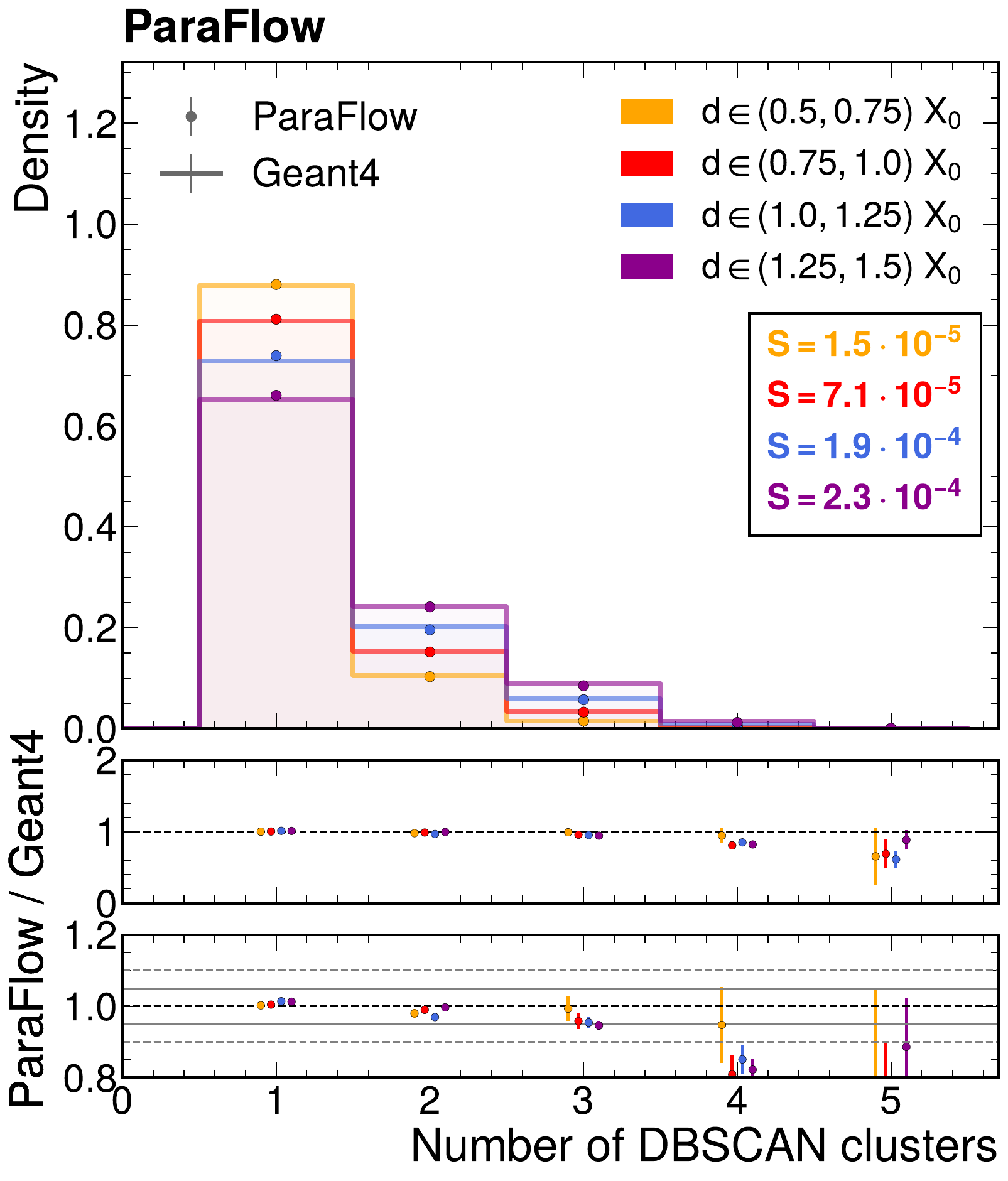}
        \end{minipage}
        \begin{minipage} {0.45 \textwidth}
            \centering
            \includegraphics[width = \textwidth]{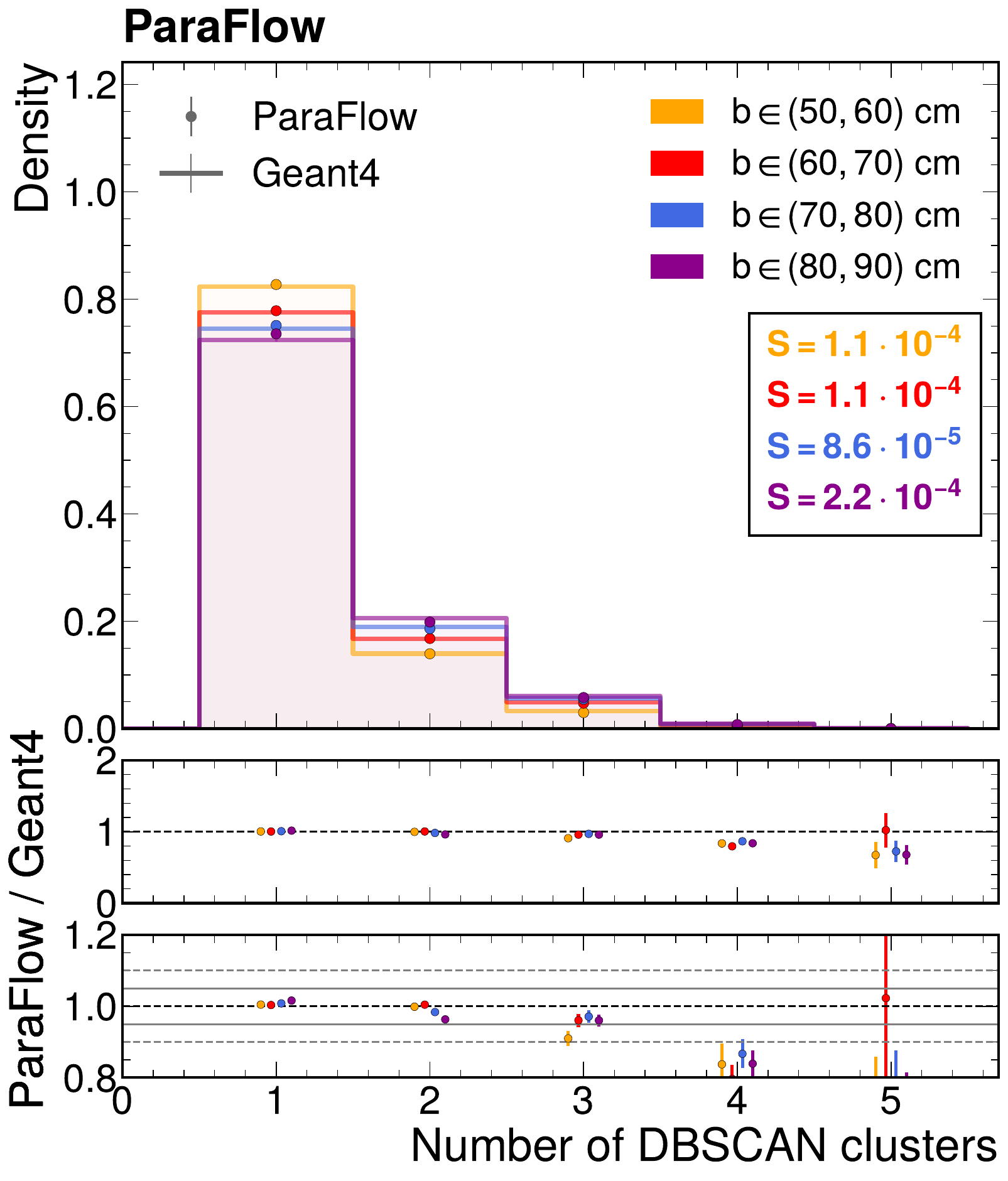}
        \end{minipage}
        \caption{Distributions of the number of energy clusters, as found by the DBSCAN algorithm, subdivided according to the values of the material parameters: distributions for varying material budget $d$ (left) and for varying material distance $b$ (right).
        The plots below show the ratio of ParaFlow over \textsc{Geant4} on two different $y$-axis scales.
        }
        \label{fig:histograms_num_clusters}
    \end{figure}

Figures~\ref{fig:histograms_shower_width}~--~\ref{fig:histograms_num_clusters} show the performance of ParaFlow in describing the material dependence of these sensitive observables:

\begin{itemize}
\item Fig.~\ref{fig:histograms_shower_width} shows the distributions of the shower width. 
    It is evident that the shower width depends on the characteristics of the material upstream of the calorimeter.
    An increase in material thickness $d$ leads to larger shower widths, as expected for a larger conversion rate.
    Larger shower widths are also expected when conversions occur earlier, i.e., with a larger distance $b$ from the calorimeter.
    These trends are clearly visible in the \textsc{Geant4} simulation, and well reproduced in the shower images generated with ParaFlow.
\item Fig.~\ref{fig:histograms_r9} shows the distribution of $R_9$. The distributions show the expected trends, i.e., lower average values and stronger tails towards low values of $R_9$ for a larger material budget (larger values of $d$) and a larger distance between the conversion vertices and the calorimeter~($b$). Again, ParaFlow reproduces these trends and closely follows the distributions from \textsc{Geant4}.
    \item Fig.~\ref{fig:histograms_num_clusters} shows the distributions of the number of energy clusters.
    We observe the expected dependence: with increasing conversion rates and for earlier conversions, the number of detected clusters tends to grow. ParaFlow reproduces these dependencies properly.
    In particular, the share of one-cluster images (unconverted photons) and two-cluster images (converted photons) is correctly reproduced at the percent level.
    For larger numbers of clusters, however, we find an underestimation by ParaFlow in comparison to the \textsc{Geant4}-simulated calorimeter images. This is possibly due to the under-representation of shower images with very large cluster counts ($\geq 4$) in the training set.
\end{itemize}

In general, the calorimeter images generated with ParaFlow agree with the predictions from \textsc{Geant4} within better than 5\% in the bulk of the distributions and with separation powers of $\mathcal{O} (10^{-4})$.
The agreement is at the same level as observed for the inclusive distribution of material-insensitive observables (Fig.~\ref{fig:insensitive_histograms}), with larger differences observed in the tails of some distributions.
However, no systematic trends are observed that would indicate that ParaFlow's performance depends on the configuration of the upstream material that it is parameterized in.

\begin{figure}[p]
    \centering
    \includegraphics[width=\linewidth]{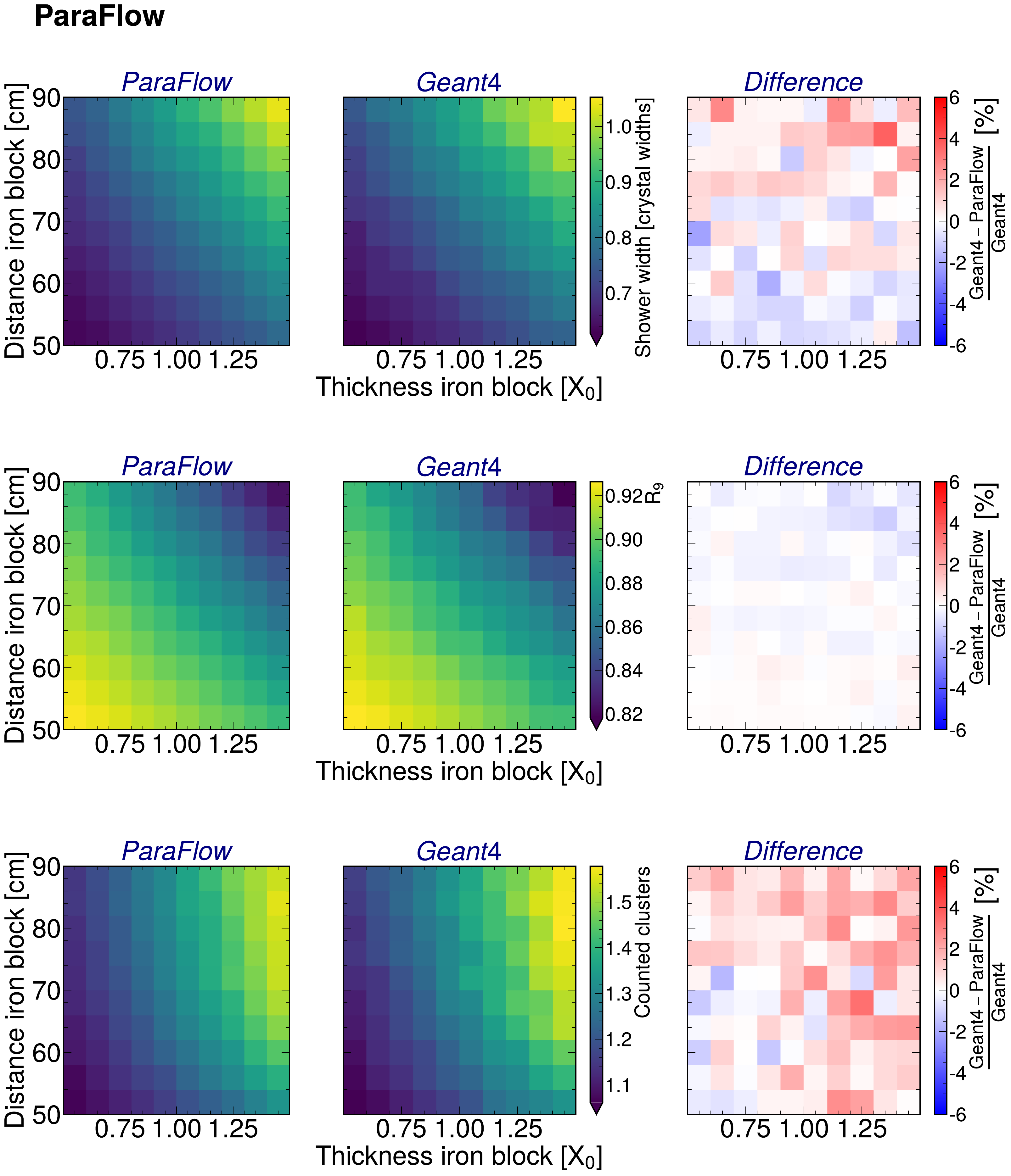}
    \caption{
    Mean value of the shower width (top row), the $R_9$ observable (middle row), and the number of clusters (bottom row) as a function the upstream material budget (thickness of the iron block, $d$, in units of $X_0$) and of the distance between the upstream material from the calorimeter (distance of the iron block, $b$, in units of $\mathrm{cm}$), shown for ParaFlow (left column) and for \textsc{Geant4} (middle column). Their relative difference (right column) is also shown.
    }
    \label{fig:2d_histograms}
\end{figure}

Figure~\ref{fig:2d_histograms} shows the mean values of these three observables (shower width, $R_9$, number of clusters), now as a function of both parameters of the upstream material, $d$ and $b$.
We observe the expected trends of the shower shape observables, i.e., wider showers, lower values of $R_9$ and more clusters, the more material is present upstream of the calorimeter and the further away it is from the calorimeter.
ParaFlow reproduces these trends very well.
The relative differences in the mean values are below $6\%$ for all of these shower shapes. For the $R_9$ variable, the deviations are even smaller, which we attribute to the fact that the distribution of $R_9$ is generally narrower (cf.~Fig.~\ref{fig:histograms_r9}).
Figure~\ref{fig:2d_histograms} hence provides further evidence that ParaFlow is able to capture the dependence of the calorimeter images on the amount and position of the upstream material.

\FloatBarrier

\section{Conclusions}

We have proposed ParaFlow, a fast calorimeter simulation that is parameterized as a function of the particle detector's configuration.
While the concept is independent of the specific generative model, we studied it based on CaloFlow~\cite{Krause:2021ilc}, which uses normalizing flows as the generative model.
We studied ParaFlow for a CMS-like toy calorimeter and for the example case of varying configurations of the upstream material.
As a simplified model for this material, we used an iron block that is characterized by its thickness and its distance from the calorimeter.
We observed the expected changes in photon calorimeter signatures in \textsc{Geant4} simulations, which we then used to train ParaFlow.
We evaluated the performance of ParaFlow by comparing its predictions for photon shower shapes to the expectation from \textsc{Geant4} simulations, as a function of the material configurations.
For the comparison, we chose a set of shower shape observables that is particularly sensitive to the properties of photon conversions to electron-positron pairs in the upstream material.

In general, we observe that ParaFlow describes the \textsc{Geant4} samples well, with an agreement of better than 5\% in the bulk of the distributions for variations of the amount of upstream material, as well as for the variations of its distance from the calorimeter.
Larger deviations are only observed in the tails of the distributions, which are not strongly represented in the training dataset.
Overall, we observe a similar separation power of $\mathcal{O}(10^{-4})$ for these shower shape observables.
We investigated the dependence of the means of the shower shape observables when both, material budget and distance, were varied simultaneously.
We find again very good agreement of the ParaFlow predictions with the expectations from the \textsc{Geant4} simulations.
We conclude that ParaFlow is indeed able to a learn a representation of calorimeter signatures as a function of properties of the upstream material in a single model.

This work provides evidence that fast calorimeter simulations can be used to learn a representation of variable configurations of the full particle detector.
This is not possible in \textsc{Geant4}, as it requires a separate simulation for each single configuration.
In future studies, ParaFlow should be extended to parameterizations of more complex variations in the detector configurations.
This may open the door to important applications:
it would allow for an efficient and more fine-grained evaluation of systematic uncertainties that are related to uncertainties in the detector configurations, such as uncertainties in the overall material budget or material distribution; it may also allow to infer parameters of detector configurations directly from data and hence improve the understanding of the detector in particle physics experiments; it may also be used as a differentiable surrogate model in gradient-based optimizations of detector designs.

\subsubsection*{Acknowledgements}

This research was supported by the Deutsche Forschungsgemeinschaft (DFG) under grants 400140256 - GRK~2497 (The physics of the heaviest particles at the LHC, JE and FM) and 686709 - ER~866/1-1 (Heisenberg Programme, JE), and by the Studienstiftung des deutschen Volkes (FM).

\phantomsection
\addcontentsline{toc}{section}{References}
\bibliographystyle{JHEP_jls}
\bibliography{main.bib}

\providecommand{\href}[2]{#2}\begingroup\raggedright\begin{thebibliography}{10}

\bibitem{GEANT4:2002zbu}
S.~Agostinelli et~al., \emph{{\textsc{Geant4} -- a simulation toolkit}}, \href{http://dx.doi.org/10.1016/S0168-9002(03)01368-8}{\emph{Nucl. Instrum. Meth. A} {\bfseries 506} (2003) 250}.

\bibitem{CERN-LHCC-2020-015}
{\scshape ATLAS} collaboration, \emph{{ATLAS HL-LHC Computing Conceptual Design Report}},  2020, CERN-LHCC-2020-015, LHCC-G-178, \url{https://cds.cern.ch/record/2729668}.

\bibitem{ATLAS:1300517}
{ATLAS collaboration}, M.~Beckingham, M.~Duehrssen, E.~Schmidt, M.~Shapiro, M.~Venturi et~al., \emph{{The simulation principle and performance of the ATLAS fast calorimeter simulation FastCaloSim}},  2010, ATL-PHYS-PUB-2010-013 \url{https://cds.cern.ch/record/1300517}.

\bibitem{ATL-SOFT-PUB-2014-001}
{ATLAS collaboration}, \emph{{Performance of the Fast ATLAS Tracking Simulation (FATRAS) and the ATLAS Fast Calorimeter Simulation (FastCaloSim) with single particles}},  2014, ATL-SOFT-PUB-2014-001, \url{https://cds.cern.ch/record/1669341}.

\bibitem{ATLAS:2021pzo}
{\scshape ATLAS} collaboration, G.~Aad et~al., \emph{{AtlFast3: The Next Generation of Fast Simulation in ATLAS}}, \href{http://dx.doi.org/10.1007/s41781-021-00079-7}{\emph{Comput. Softw. Big Sci.} {\bfseries 6} (2022) 7}, [\href{https://arxiv.org/abs/2109.02551}{{\ttfamily 2109.02551}}].

\bibitem{Abdullin:2011zz}
{\scshape CMS} collaboration, S.~Abdullin, P.~Azzi, F.~Beaudette, P.~Janot and A.~Perrotta, \emph{{The Fast Simulation of the CMS Detector at LHC}}, \href{http://dx.doi.org/10.1088/1742-6596/331/3/032049}{\emph{J. Phys. Conf. Ser.} {\bfseries 331} (2011) 032049}.

\bibitem{Hildreth:2297284}
{\scshape CMS} collaboration, M.~Hildreth, V.~N. Ivanchenko and D.~J. Lange, \emph{{Upgrades for the CMS simulation}}, \href{http://dx.doi.org/10.1088/1742-6596/898/4/042040}{\emph{J. Phys. Conf. Ser.} {\bfseries 898} (2017) 042040}.

\bibitem{Paganini:2017dwg}
M.~Paganini, L.~de~Oliveira and B.~Nachman, \emph{{CaloGAN: Simulating 3D high energy particle showers in multilayer electromagnetic calorimeters with generative adversarial networks}}, \href{http://dx.doi.org/10.1103/PhysRevD.97.014021}{\emph{Phys. Rev. D} {\bfseries 97} (2018) 014021}, [\href{https://arxiv.org/abs/1712.10321}{{\ttfamily 1712.10321}}].

\bibitem{Hashemi:2023rgo}
B.~Hashemi and C.~Krause, \emph{{Deep generative models for detector signature simulation: A taxonomic review}}, \href{http://dx.doi.org/10.1016/j.revip.2024.100092}{\emph{Rev. Phys.} {\bfseries 12} (2024) 100092}, [\href{https://arxiv.org/abs/2312.09597}{{\ttfamily 2312.09597}}].

\bibitem{Ahmad:2024dql}
F.~Y. Ahmad, V.~Venkataswamy and G.~Fox, \emph{{A Comprehensive Evaluation of Generative Models in Calorimeter Shower Simulation}},  \href{https://arxiv.org/abs/2406.12898}{{\ttfamily 2406.12898}}.

\bibitem{Barbetti:2023bvi}
M.~Barbetti, \emph{{Lamarr: LHCb ultra-fast simulation based on machine learning models deployed within Gauss}},  {Proceedings of the 21st International Workshop on Advanced Computing and Analysis Techniques in Physics Research (ACAT 2022)}, \href{https://arxiv.org/abs/2303.11428}{{\ttfamily 2303.11428}}.

\bibitem{ATLAS:2022jhk}
{\scshape ATLAS} collaboration, G.~Aad et~al., \emph{{Deep Generative Models for Fast Photon Shower Simulation in ATLAS}}, \href{http://dx.doi.org/10.1007/s41781-023-00106-9}{\emph{Comput. Softw. Big Sci.} {\bfseries 8} (2024) 7}, [\href{https://arxiv.org/abs/2210.06204}{{\ttfamily 2210.06204}}].

\bibitem{deja2020endtoendsinkhornautoencodernoise}
K.~Deja, J.~Dubiński, P.~Nowak, S.~Wenzel and T.~Trzciński, \emph{{End-to-end Sinkhorn Autoencoder with Noise Generator}},  \href{https://arxiv.org/abs/2006.06704}{{\ttfamily 2006.06704}}.

\bibitem{Buhmann:2020pmy}
E.~Buhmann, S.~Diefenbacher, E.~Eren, F.~Gaede, G.~Kasieczka, A.~Korol et~al., \emph{{Getting High: High Fidelity Simulation of High Granularity Calorimeters with High Speed}}, \href{http://dx.doi.org/10.1007/s41781-021-00056-0}{\emph{Comput. Softw. Big Sci.} {\bfseries 5} (2021) 13}, [\href{https://arxiv.org/abs/2005.05334}{{\ttfamily 2005.05334}}].

\bibitem{Buhmann:2021lxj}
E.~Buhmann, S.~Diefenbacher, E.~Eren, F.~Gaede, G.~Kasieczka, A.~Korol et~al., \emph{{Decoding Photons: Physics in the Latent Space of a BIB-AE Generative Network}}, \href{http://dx.doi.org/10.1051/epjconf/202125103003}{\emph{EPJ Web Conf.} {\bfseries 251} (2021) 03003}, [\href{https://arxiv.org/abs/2102.12491}{{\ttfamily 2102.12491}}].

\bibitem{Buhmann:2021caf}
E.~Buhmann, S.~Diefenbacher, D.~Hundhausen, G.~Kasieczka, W.~Korcari, E.~Eren et~al., \emph{{Hadrons, better, faster, stronger}}, \href{http://dx.doi.org/10.1088/2632-2153/ac7848}{\emph{Mach. Learn. Sci. Tech.} {\bfseries 3} (2022) 025014}, [\href{https://arxiv.org/abs/2112.09709}{{\ttfamily 2112.09709}}].

\bibitem{Diefenbacher:2023prl}
S.~Diefenbacher, E.~Eren, F.~Gaede, G.~Kasieczka, A.~Korol, K.~Kr\"uger et~al., \emph{{New angles on fast calorimeter shower simulation}}, \href{http://dx.doi.org/10.1088/2632-2153/acefa9}{\emph{Mach. Learn. Sci. Tech.} {\bfseries 4} (2023) 035044}, [\href{https://arxiv.org/abs/2303.18150}{{\ttfamily 2303.18150}}].

\bibitem{Hariri:2021clz}
A.~Hariri, D.~Dyachkova and S.~Gleyzer, \emph{{Graph Generative Models for Fast Detector Simulations in High Energy Physics}},  \href{https://arxiv.org/abs/2104.01725}{{\ttfamily 2104.01725}}.

\bibitem{AbhishekAbhishek:2022wby}
A.~Abhishek, E.~Drechsler, W.~Fedorko and B.~Stelzer, \emph{{CaloDVAE : Discrete Variational Autoencoders for Fast Calorimeter Shower Simulation}},  {Proceedings of the 35th Conference on Neural Information Processing Systems (NeurIPS 2021) Workshop on Machine Learning and the Physical Sciences}, \href{https://arxiv.org/abs/2210.07430}{{\ttfamily 2210.07430}}.

\bibitem{Cresswell:2022tof}
J.~C. Cresswell, B.~L. Ross, G.~Loaiza-Ganem, H.~Reyes-Gonzalez, M.~Letizia and A.~L. Caterini, \emph{{CaloMan: Fast generation of calorimeter showers with density estimation on learned manifolds}},  {Proceedings of the 36th Conference on Neural Information Processing Systems (NeurIPS 2022) Workshop on Machine Learning and the Physical Sciences}, \href{https://arxiv.org/abs/2211.15380}{{\ttfamily 2211.15380}}.

\bibitem{Liu:2024kvv}
Q.~Liu, C.~Shimmin, X.~Liu, E.~Shlizerman, S.~Li and S.-C. Hsu, \emph{{Calo-VQ: Vector-Quantized Two-Stage Generative Model in Calorimeter Simulation}},  \href{https://arxiv.org/abs/2405.06605}{{\ttfamily 2405.06605}}.

\bibitem{deOliveira:2017pjk}
L.~de~Oliveira, M.~Paganini and B.~Nachman, \emph{{Learning Particle Physics by Example: Location-Aware Generative Adversarial Networks for Physics Synthesis}}, \href{http://dx.doi.org/10.1007/s41781-017-0004-6}{\emph{Comput. Softw. Big Sci.} {\bfseries 1} (2017) 4}, [\href{https://arxiv.org/abs/1701.05927}{{\ttfamily 1701.05927}}].

\bibitem{deOliveira:2017rwa}
L.~de~Oliveira, M.~Paganini and B.~Nachman, \emph{{Controlling Physical Attributes in GAN-Accelerated Simulation of Electromagnetic Calorimeters}}, \href{http://dx.doi.org/10.1088/1742-6596/1085/4/042017}{\emph{J. Phys. Conf. Ser.} {\bfseries 1085} (2018) 042017}, [\href{https://arxiv.org/abs/1711.08813}{{\ttfamily 1711.08813}}].

\bibitem{Paganini:2017hrr}
M.~Paganini, L.~de~Oliveira and B.~Nachman, \emph{{Accelerating Science with Generative Adversarial Networks: An Application to 3D Particle Showers in Multilayer Calorimeters}}, \href{http://dx.doi.org/10.1103/PhysRevLett.120.042003}{\emph{Phys. Rev. Lett.} {\bfseries 120} (2018) 042003}, [\href{https://arxiv.org/abs/1705.02355}{{\ttfamily 1705.02355}}].

\bibitem{Khattak:2021ndw}
G.~R. Khattak, S.~Vallecorsa, F.~Carminati and G.~M. Khan, \emph{{Fast simulation of a high granularity calorimeter by generative adversarial networks}}, \href{http://dx.doi.org/10.1140/epjc/s10052-022-10258-4}{\emph{Eur. Phys. J. C} {\bfseries 82} (2022) 386}, [\href{https://arxiv.org/abs/2109.07388}{{\ttfamily 2109.07388}}].

\bibitem{8451587}
G.~R. Khattak, S.~Vallecorsa and F.~Carminati, \emph{{Three Dimensional Energy Parametrized Generative Adversarial Networks for Electromagnetic Shower Simulation}}, \href{http://dx.doi.org/10.1109/ICIP.2018.8451587}{\emph{Proceedings of the 25th IEEE International Conference on Image Processing (ICIP 2018)} (2018) 3913}.

\bibitem{Vallecorsa:2019ked}
S.~Vallecorsa, F.~Carminati and G.~Khattak, \emph{{3D convolutional GAN for fast simulation}}, \href{http://dx.doi.org/10.1051/epjconf/201921402010}{\emph{EPJ Web Conf.} {\bfseries 214} (2019) 02010}.

\bibitem{Belayneh:2019vyx}
D.~Belayneh et~al., \emph{{Calorimetry with deep learning: particle simulation and reconstruction for collider physics}}, \href{http://dx.doi.org/10.1140/epjc/s10052-020-8251-9}{\emph{Eur. Phys. J. C} {\bfseries 80} (2020) 688}, [\href{https://arxiv.org/abs/1912.06794}{{\ttfamily 1912.06794}}].

\bibitem{Musella:2018rdi}
P.~Musella and F.~Pandolfi, \emph{{Fast and Accurate Simulation of Particle Detectors Using Generative Adversarial Networks}}, \href{http://dx.doi.org/10.1007/s41781-018-0015-y}{\emph{Comput. Softw. Big Sci.} {\bfseries 2} (2018) 8}, [\href{https://arxiv.org/abs/1805.00850}{{\ttfamily 1805.00850}}].

\bibitem{Chekalina:2018hxi}
V.~Chekalina, E.~Orlova, F.~Ratnikov, D.~Ulyanov, A.~Ustyuzhanin and E.~Zakharov, \emph{{Generative Models for Fast Calorimeter Simulation: the LHCb case}}, \href{http://dx.doi.org/10.1051/epjconf/201921402034}{\emph{EPJ Web Conf.} {\bfseries 214} (2019) 02034}, [\href{https://arxiv.org/abs/1812.01319}{{\ttfamily 1812.01319}}].

\bibitem{Diefenbacher:2020rna}
S.~Diefenbacher, E.~Eren, G.~Kasieczka, A.~Korol, B.~Nachman and D.~Shih, \emph{{DCTRGAN: improving the precision of generative models with reweighting}}, \href{http://dx.doi.org/10.1088/1748-0221/15/11/P11004}{\emph{JINST} {\bfseries 15} (2020) P11004}, [\href{https://arxiv.org/abs/2009.03796}{{\ttfamily 2009.03796}}].

\bibitem{Jaruskova:2023cke}
K.~Jaruskova and S.~Vallecorsa, \emph{{Ensemble Models for Calorimeter Simulations}}, \href{http://dx.doi.org/10.1088/1742-6596/2438/1/012080}{\emph{J. Phys. Conf. Ser.} {\bfseries 2438} (2023) 012080}.

\bibitem{FaucciGiannelli:2023fow}
M.~Faucci~Giannelli and R.~Zhang, \emph{{CaloShowerGAN, a generative adversarial network model for fast calorimeter shower simulation}}, \href{http://dx.doi.org/10.1140/epjp/s13360-024-05397-4}{\emph{Eur. Phys. J. Plus} {\bfseries 139} (2024) 597}, [\href{https://arxiv.org/abs/2309.06515}{{\ttfamily 2309.06515}}].

\bibitem{Erdmann:2018jxd}
M.~Erdmann, J.~Glombitza and T.~Quast, \emph{{Precise Simulation of Electromagnetic Calorimeter Showers Using a Wasserstein Generative Adversarial Network}}, \href{http://dx.doi.org/10.1007/s41781-018-0019-7}{\emph{Comput. Softw. Big Sci.} {\bfseries 3} (2019) 4}, [\href{https://arxiv.org/abs/1807.01954}{{\ttfamily 1807.01954}}].

\bibitem{Carminati:2018khv}
F.~Carminati, A.~Gheata, G.~Khattak, P.~Mendez~Lorenzo, S.~Sharan and S.~Vallecorsa, \emph{{Three dimensional Generative Adversarial Networks for fast simulation}}, \href{http://dx.doi.org/10.1088/1742-6596/1085/3/032016}{\emph{J. Phys. Conf. Ser.} {\bfseries 1085} (2018) 032016}.

\bibitem{Erdmann:2023ngr}
J.~Erdmann, A.~van~der Graaf, F.~Mausolf and O.~Nackenhorst, \emph{{SR-GAN for SR-gamma: super resolution of photon calorimeter images at collider experiments}}, \href{http://dx.doi.org/10.1140/epjc/s10052-023-12178-3}{\emph{Eur. Phys. J. C} {\bfseries 83} (2023) 1001}, [\href{https://arxiv.org/abs/2308.09025}{{\ttfamily 2308.09025}}].

\bibitem{Krause:2021ilc}
C.~Krause and D.~Shih, \emph{{Fast and accurate simulations of calorimeter showers with normalizing flows}}, \href{http://dx.doi.org/10.1103/PhysRevD.107.113003}{\emph{Phys. Rev. D} {\bfseries 107} (2023) 113003}, [\href{https://arxiv.org/abs/2106.05285}{{\ttfamily 2106.05285}}].

\bibitem{Krause:2021wez}
C.~Krause and D.~Shih, \emph{{Accelerating accurate simulations of calorimeter showers with normalizing flows and probability density distillation}}, \href{http://dx.doi.org/10.1103/PhysRevD.107.113004}{\emph{Phys. Rev. D} {\bfseries 107} (2023) 113004}, [\href{https://arxiv.org/abs/2110.11377}{{\ttfamily 2110.11377}}].

\bibitem{Krause:2022jna}
C.~Krause, I.~Pang and D.~Shih, \emph{{CaloFlow for CaloChallenge dataset 1}}, \href{http://dx.doi.org/10.21468/SciPostPhys.16.5.126}{\emph{SciPost Phys.} {\bfseries 16} (2024) 126}, [\href{https://arxiv.org/abs/2210.14245}{{\ttfamily 2210.14245}}].

\bibitem{Pang:2023wfx}
I.~Pang, D.~Shih and J.~A. Raine, \emph{{Calorimeter shower superresolution}}, \href{http://dx.doi.org/10.1103/PhysRevD.109.092009}{\emph{Phys. Rev. D} {\bfseries 109} (2024) 092009}, [\href{https://arxiv.org/abs/2308.11700}{{\ttfamily 2308.11700}}].

\bibitem{CaloPointFlowI}
S.~Schnake, D.~Kr\"uker and K.~Borras, \emph{{Generating Calorimeter Showers as Point Clouds}},  Paper at Workshop Machiner Learning and the Physical Sciences at the 37th conference on Neural Information Processing Systems (NeurIPS 2022), \url{https://ml4physicalsciences.github.io/2022/files/NeurIPS_ML4PS_2022_77.pdf}.

\bibitem{Schnake:2024mip}
S.~Schnake, D.~Kr\"ucker and K.~Borras, \emph{{CaloPointFlow II Generating Calorimeter Showers as Point Clouds}},  \href{https://arxiv.org/abs/2403.15782}{{\ttfamily 2403.15782}}.

\bibitem{Buss:2024orz}
T.~Buss, F.~Gaede, G.~Kasieczka, C.~Krause and D.~Shih, \emph{{Convolutional L2LFlows: generating accurate showers in highly granular calorimeters using convolutional normalizing flows}}, \href{http://dx.doi.org/10.1088/1748-0221/19/09/P09003}{\emph{JINST} {\bfseries 19} (2024) P09003}, [\href{https://arxiv.org/abs/2405.20407}{{\ttfamily 2405.20407}}].

\bibitem{Dreyer:2024bhs}
E.~Dreyer, E.~Gross, D.~Kobylianskii, V.~Mikuni, B.~Nachman and N.~Soybelman, \emph{{Automated Approach to Accurate, Precise, and Fast Detector Simulation and Reconstruction}}, \href{http://dx.doi.org/10.1103/PhysRevLett.133.211902}{\emph{Phys. Rev. Lett.} {\bfseries 133} (2024) 211902}, [\href{https://arxiv.org/abs/2406.01620}{{\ttfamily 2406.01620}}].

\bibitem{Ernst:2023qvn}
F.~Ernst, L.~Favaro, C.~Krause, T.~Plehn and D.~Shih, \emph{{Normalizing flows for high-dimensional detector simulations}}, \href{http://dx.doi.org/10.21468/SciPostPhys.18.3.081}{\emph{SciPost Phys.} {\bfseries 18} (2025) 081}, [\href{https://arxiv.org/abs/2312.09290}{{\ttfamily 2312.09290}}].

\bibitem{Favaro:2024rle}
L.~Favaro, A.~Ore, S.~P. Schweitzer and T.~Plehn, \emph{{CaloDREAM -- Detector response emulation via attentive flow matching}}, \href{http://dx.doi.org/10.21468/SciPostPhys.18.3.088}{\emph{SciPost Phys.} {\bfseries 18} (2025) 088}, [\href{https://arxiv.org/abs/2405.09629}{{\ttfamily 2405.09629}}].

\bibitem{Mikuni:2022xry}
V.~Mikuni and B.~Nachman, \emph{{Score-based generative models for calorimeter shower simulation}}, \href{http://dx.doi.org/10.1103/PhysRevD.106.092009}{\emph{Phys. Rev. D} {\bfseries 106} (2022) 092009}, [\href{https://arxiv.org/abs/2206.11898}{{\ttfamily 2206.11898}}].

\bibitem{Mikuni:2023tqg}
V.~Mikuni and B.~Nachman, \emph{{CaloScore v2: single-shot calorimeter shower simulation with diffusion models}}, \href{http://dx.doi.org/10.1088/1748-0221/19/02/P02001}{\emph{JINST} {\bfseries 19} (2024) P02001}, [\href{https://arxiv.org/abs/2308.03847}{{\ttfamily 2308.03847}}].

\bibitem{Buhmann:2023bwk}
E.~Buhmann, S.~Diefenbacher, E.~Eren, F.~Gaede, G.~Kasieczka, A.~Korol et~al., \emph{{CaloClouds: fast geometry-independent highly-granular calorimeter simulation}}, \href{http://dx.doi.org/10.1088/1748-0221/18/11/P11025}{\emph{JINST} {\bfseries 18} (2023) P11025}, [\href{https://arxiv.org/abs/2305.04847}{{\ttfamily 2305.04847}}].

\bibitem{Buhmann:2023kdg}
E.~Buhmann, F.~Gaede, G.~Kasieczka, A.~Korol, W.~Korcari, K.~Kr\"uger et~al., \emph{{CaloClouds~II: ultra-fast geometry-independent highly-granular calorimeter simulation}}, \href{http://dx.doi.org/10.1088/1748-0221/19/04/P04020}{\emph{JINST} {\bfseries 19} (2024) P04020}, [\href{https://arxiv.org/abs/2309.05704}{{\ttfamily 2309.05704}}].

\bibitem{Diefenbacher:2023flw}
S.~Diefenbacher, V.~Mikuni and B.~Nachman, \emph{{Refining Fast Calorimeter Simulations with a Schr\"odinger Bridge}},  \href{https://arxiv.org/abs/2308.12339}{{\ttfamily 2308.12339}}.

\bibitem{Amram:2023onf}
O.~Amram and K.~Pedro, \emph{{Denoising diffusion models with geometry adaptation for high fidelity calorimeter simulation}}, \href{http://dx.doi.org/10.1103/PhysRevD.108.072014}{\emph{Phys. Rev. D} {\bfseries 108} (2023) 072014}, [\href{https://arxiv.org/abs/2308.03876}{{\ttfamily 2308.03876}}].

\bibitem{Acosta:2023zik}
F.~T. Acosta, V.~Mikuni, B.~Nachman, M.~Arratia, B.~Karki, R.~Milton et~al., \emph{{Comparison of point cloud and image-based models for calorimeter fast simulation}}, \href{http://dx.doi.org/10.1088/1748-0221/19/05/P05003}{\emph{JINST} {\bfseries 19} (2024) P05003}, [\href{https://arxiv.org/abs/2307.04780}{{\ttfamily 2307.04780}}].

\bibitem{Kobylianskii:2024ijw}
D.~Kobylianskii, N.~Soybelman, E.~Dreyer and E.~Gross, \emph{{Graph-based diffusion model for fast shower generation in calorimeters with irregular geometry}}, \href{http://dx.doi.org/10.1103/PhysRevD.110.072003}{\emph{Phys. Rev. D} {\bfseries 110} (2024) 072003}, [\href{https://arxiv.org/abs/2402.11575}{{\ttfamily 2402.11575}}].

\bibitem{Kobylianskii:2024sup}
D.~Kobylianskii, N.~Soybelman, N.~Kakati, E.~Dreyer, B.~Nachman and E.~Gross, \emph{{Advancing set-conditional set generation: Diffusion models for fast simulation of reconstructed particles}}, \href{http://dx.doi.org/10.1103/PhysRevD.110.092013}{\emph{Phys. Rev. D} {\bfseries 110} (2024) 092013}, [\href{https://arxiv.org/abs/2405.10106}{{\ttfamily 2405.10106}}].

\bibitem{madula_mikuni}
T.~Madula and V.~M. Mikuni, \emph{{CaloLatent: Score-based Generative Modelling in the Latent Space for Calorimeter Shower Generation}},  Paper at Workshop Machiner Learning and the Physical Sciences at the 38th conference on Neural Information Processing Systems (NeurIPS 2023), \url{https://ml4physicalsciences.github.io/2023/files/NeurIPS_ML4PS_2023_19.pdf}.

\bibitem{Lu:2020npg}
Y.~Lu, J.~Collado, D.~Whiteson and P.~Baldi, \emph{{Sparse autoregressive models for scalable generation of sparse images in particle physics}}, \href{http://dx.doi.org/10.1103/PhysRevD.103.036012}{\emph{Phys. Rev. D} {\bfseries 103} (2021) 036012}, [\href{https://arxiv.org/abs/2009.14017}{{\ttfamily 2009.14017}}].

\bibitem{Liu:2022dem}
J.~Liu, A.~Ghosh, D.~Smith, P.~Baldi and D.~Whiteson, \emph{{Geometry-aware Autoregressive Models for Calorimeter Shower Simulations}},  {Proceedings of the 36th Conference on Neural Information Processing Systems (NeurIPS 2022) Workshop on Machine Learning and the Physical Sciences}, \href{https://arxiv.org/abs/2212.08233}{{\ttfamily 2212.08233}}.

\bibitem{Liu:2023lnn}
J.~Liu, A.~Ghosh, D.~Smith, P.~Baldi and D.~Whiteson, \emph{{Generalizing to new geometries with Geometry-Aware Autoregressive Models (GAAMs) for fast calorimeter simulation}}, \href{http://dx.doi.org/10.1088/1748-0221/18/11/P11003}{\emph{JINST} {\bfseries 18} (2023) P11003}, [\href{https://arxiv.org/abs/2305.11531}{{\ttfamily 2305.11531}}].

\bibitem{Diefenbacher:2023vsw}
S.~Diefenbacher, E.~Eren, F.~Gaede, G.~Kasieczka, C.~Krause, I.~Shekhzadeh et~al., \emph{{L2LFlows: generating high-fidelity 3D calorimeter images}}, \href{http://dx.doi.org/10.1088/1748-0221/18/10/P10017}{\emph{JINST} {\bfseries 18} (2023) P10017}, [\href{https://arxiv.org/abs/2302.11594}{{\ttfamily 2302.11594}}].

\bibitem{Buckley:2023daw}
M.~R. Buckley, C.~Krause, I.~Pang and D.~Shih, \emph{{Inductive simulation of calorimeter showers with normalizing flows}}, \href{http://dx.doi.org/10.1103/PhysRevD.109.033006}{\emph{Phys. Rev. D} {\bfseries 109} (2024) 033006}, [\href{https://arxiv.org/abs/2305.11934}{{\ttfamily 2305.11934}}].

\bibitem{Krause:2024avx}
C.~Krause, M.~Faucci~Giannelli, G.~Kasieczka, B.~Nachman, D.~Salamani, D.~Shih et~al., \emph{{CaloChallenge 2022: A Community Challenge for Fast Calorimeter Simulation}},  \href{https://arxiv.org/abs/2410.21611}{{\ttfamily 2410.21611}}.

\bibitem{Smith:2024lxz}
D.~Smith, A.~Ghosh, J.~Liu, P.~Baldi and D.~Whiteson, \emph{{Fast multi-geometry calorimeter simulation with conditional self-attention variational autoencoders}},  \href{https://arxiv.org/abs/2411.05996}{{\ttfamily 2411.05996}}.

\bibitem{Raikwar:2024peb}
P.~Raikwar, R.~Cardoso, N.~Chernyavskaya, K.~Jaruskova, W.~Pokorski, D.~Salamani et~al., \emph{{Transformers for Generalized Fast Shower Simulation}}, \href{http://dx.doi.org/10.1051/epjconf/202429509039}{\emph{EPJ Web Conf.} {\bfseries 295} (2024) 09039}.

\bibitem{Salamani:2023ttx}
D.~Salamani, A.~Zaborowska and W.~Pokorski, \emph{{MetaHEP: Meta learning for fast shower simulation of high energy physics experiments}}, \href{http://dx.doi.org/10.1016/j.physletb.2023.138079}{\emph{Phys. Lett. B} {\bfseries 844} (2023) 138079}.

\bibitem{ATLAS:2023owm}
{\scshape ATLAS} collaboration, G.~Aad et~al., \emph{{Measurement of the Higgs boson mass with $H\rightarrow\gamma\gamma$ decays in 140~fb$^{-1}$ of $\sqrt{s} = 13$ TeV pp collisions with the ATLAS detector}}, \href{http://dx.doi.org/10.1016/j.physletb.2023.138315}{\emph{Phys. Lett. B} {\bfseries 847} (2023) 138315}, [\href{https://arxiv.org/abs/2308.07216}{{\ttfamily 2308.07216}}].

\bibitem{CMS:2020xrn}
{\scshape CMS} collaboration, A.~M. Sirunyan et~al., \emph{{A measurement of the Higgs boson mass in the diphoton decay channel}}, \href{http://dx.doi.org/10.1016/j.physletb.2020.135425}{\emph{Phys. Lett. B} {\bfseries 805} (2020) 135425}, [\href{https://arxiv.org/abs/2002.06398}{{\ttfamily 2002.06398}}].

\bibitem{MODE:2021yid}
{\scshape MODE} collaboration, A.~G. Baydin et~al., \emph{{Toward Machine Learning Optimization of Experimental Design}}, \href{http://dx.doi.org/10.1080/10619127.2021.1881364}{\emph{Nuclear Physics News} {\bfseries 31} (2021) 25}.

\bibitem{MODE:2022znx}
{\scshape MODE} collaboration, T.~Dorigo et~al., \emph{{Toward the end-to-end optimization of particle physics instruments with differentiable programming}}, \href{http://dx.doi.org/10.1016/j.revip.2023.100085}{\emph{Reviews in Physics} {\bfseries 10} (2023) 100085}, [\href{https://arxiv.org/abs/2203.13818}{{\ttfamily 2203.13818}}].

\bibitem{Shirobokov:2020tjt}
S.~Shirobokov, V.~Belavin, M.~Kagan, A.~Ustyuzhanin and A.~G. Baydin, \emph{{Black-Box Optimization with Local Generative Surrogates}},  {Proceedings of Advances in Neural Information Processing Systems 34 (NeurIPS 2020)}, \href{https://arxiv.org/abs/2002.04632}{{\ttfamily 2002.04632}}.

\bibitem{Neubuser:2021uui}
C.~Neub\"user, J.~Kieseler and P.~Lujan, \emph{{Optimising longitudinal and lateral calorimeter granularity for software compensation in hadronic showers using deep neural networks}}, \href{http://dx.doi.org/10.1140/epjc/s10052-022-10031-7}{\emph{Eur. Phys. J.~C} {\bfseries 82} (2022) 92}, [\href{https://arxiv.org/abs/2101.08150}{{\ttfamily 2101.08150}}].

\bibitem{Aehle:2023wwi}
M.~Aehle et~al., \emph{{Progress in End-to-End Optimization of Detectors for Fundamental Physics with Differentiable Programming}},  \href{https://arxiv.org/abs/2310.05673}{{\ttfamily 2310.05673}}.

\bibitem{Aehle:2024ezu}
M.~Aehle, M.~Nov\'ak, V.~Vassilev, N.~R. Gauger, L.~Heinrich, M.~Kagan et~al., \emph{{Optimization Using Pathwise Algorithmic Derivatives of Electromagnetic Shower Simulations}},  \href{https://arxiv.org/abs/2405.07944}{{\ttfamily 2405.07944}}.

\bibitem{Strong:2023oew}
G.~C. Strong et~al., \emph{{TomOpt: differential optimisation for task- and constraint-aware design of particle detectors in the context of muon tomography}}, \href{http://dx.doi.org/10.1088/2632-2153/ad52e7}{\emph{Mach. Learn. Sci. Tech.} {\bfseries 5} (2024) 035002}, [\href{https://arxiv.org/abs/2309.14027}{{\ttfamily 2309.14027}}].

\bibitem{MODE:2025zir}
{\scshape MODE} collaboration, K.~Schmidt et~al., \emph{{End-to-End Detector Optimization with Diffusion models: A Case Study in Sampling Calorimeters}},  \href{https://arxiv.org/abs/2502.02152}{{\ttfamily 2502.02152}}.

\bibitem{Wozniak:2025ttb}
K.~A. Wo\'zniak, S.~Mulligan, J.~Kieseler, M.~Klute, F.~Fleuret and T.~Golling, \emph{{End-to-End Optimal Detector Design with Mutual Information Surrogates}},  \href{https://arxiv.org/abs/2503.14342}{{\ttfamily 2503.14342}}.

\bibitem{Tabak}
E.~G. Tabak and C.~V. Turner, \emph{{A Family of Nonparametric Density Estimation Algorithms}}, \href{http://dx.doi.org/https://doi.org/10.1002/cpa.21423}{\emph{Commun. Pure Appl. Math.} {\bfseries 66} (2013) 145}.

\bibitem{papamakarios2021normalizing}
G.~Papamakarios, E.~Nalisnick, D.~J. Rezende, S.~Mohamed and B.~Lakshminarayanan, \emph{{Normalizing Flows for Probabilistic Modeling and Inference}}, {\emph{J. Mach. Learn. Res.} {\bfseries 22} (2021) 1}, [\href{https://arxiv.org/abs/1912.02762}{{\ttfamily 1912.02762}}].

\bibitem{CMS:2008xjf}
{\scshape CMS} collaboration, S.~Chatrchyan et~al., \emph{{The CMS Experiment at the CERN LHC}}, \href{http://dx.doi.org/10.1088/1748-0221/3/08/S08004}{\emph{JINST} {\bfseries 3} (2008) S08004}.

\bibitem{CMS:2014pgm}
{\scshape CMS} collaboration, S.~Chatrchyan et~al., \emph{{Description and performance of track and primary-vertex reconstruction with the CMS tracker}}, \href{http://dx.doi.org/10.1088/1748-0221/9/10/P10009}{\emph{JINST} {\bfseries 9} (2014) P10009}, [\href{https://arxiv.org/abs/1405.6569}{{\ttfamily 1405.6569}}].

\bibitem{papamakarios2017masked}
G.~Papamakarios, T.~Pavlakou and I.~Murray, \emph{{Masked Autoregressive Flow for Density Estimation}},  {Proceedings of Advances in Neural Information Processing Systems 30 (NIPS 2017)}, \href{https://arxiv.org/abs/1705.07057}{{\ttfamily 1705.07057}}.

\bibitem{durkan2019neural}
C.~Durkan, A.~Bekasov, I.~Murray and G.~Papamakarios, \emph{{Neural Spline Flows}},  Proceedings of the 32nd Conference on Neural Information Processing Systems (NeurIPS 2019), \href{https://arxiv.org/abs/1906.04032}{{\ttfamily 1906.04032}}.

\bibitem{germain2015made}
M.~Germain, K.~Gregor, I.~Murray and H.~Larochelle, \emph{{MADE: Masked Autoencoder for Distribution Estimation}},  Proceedings of the 32nd International Conference on Machine Learning (ICML 2015), PMLR 37 (2015) 881, \href{https://arxiv.org/abs/1502.03509}{{\ttfamily 1502.03509}}.

\bibitem{he2015deepresiduallearningimage}
K.~He, X.~Zhang, S.~Ren and J.~Sun, \emph{{Deep Residual Learning for Image Recognition}},  Proceedings of the 2016 Conference on Computer Vision and Pattern Recognition (CVPR 2016), \href{https://arxiv.org/abs/1512.03385}{{\ttfamily 1512.03385}}.

\bibitem{paszke2019pytorch}
A.~Paszke, S.~Gross, F.~Massa, A.~Lerer, J.~Bradbury, G.~Chanan et~al., \emph{{PyTorch: An Imperative Style, High-Performance Deep Learning Library}},  Proceedings of the 32nd Conference on Neural Information Processing Systems (NeurIPS 2019), \href{https://arxiv.org/abs/1912.01703}{{\ttfamily 1912.01703}}.

\bibitem{rozet2022zuko}
F.~Rozet et~al., \emph{{Zuko: Normalizing flows in PyTorch}},  \url{https://pypi.org/project/zuko}.

\bibitem{Kingma:2014vow}
D.~P. Kingma and J.~Ba, \emph{{Adam: A Method for Stochastic Optimization}},  Proceedings of the 3rd International Conference on Learning Representations (ICLR 2015), \href{https://arxiv.org/abs/1412.6980}{{\ttfamily 1412.6980}}.

\bibitem{CosineAnnealing}
I.~Loshchilov and F.~Hutter, \emph{{SGDR}: {Stochastic Gradient Descent with Warm Restarts}},  Proceedings of the 5th International Conference on Learning Representations (ICLR 2017), \href{https://arxiv.org/abs/1608.03983}{{\ttfamily 1608.03983}}.

\bibitem{CMS:2015myp}
{\scshape CMS} collaboration, V.~Khachatryan et~al., \emph{{{Performance of photon reconstruction and identification with the CMS detector in proton-proton collisions at \ensuremath{\sqrt{s}} = 8 TeV}}}, \href{http://dx.doi.org/10.1088/1748-0221/10/08/P08010}{\emph{JINST} {\bfseries 10} (2015) P08010}, [\href{https://arxiv.org/abs/1502.02702}{{\ttfamily 1502.02702}}].

\bibitem{CMS:2020uim}
{\scshape CMS} collaboration, A.~M. Sirunyan et~al., \emph{{{Electron and photon reconstruction and identification with the CMS experiment at the CERN LHC}}}, \href{http://dx.doi.org/10.1088/1748-0221/16/05/P05014}{\emph{JINST} {\bfseries 16} (2021) P05014}, [\href{https://arxiv.org/abs/2012.06888}{{\ttfamily 2012.06888}}].

\bibitem{DBSCAN}
M.~Ester, H.~Kriegel, J.~Sander and X.~Xu, \emph{{A density-based algorithm for discovering clusters in large spatial databases with noise}},  Proceedings of the Second International Conference on Knowledge Discovery and Data Mining (KDD'96) 226, \url{https://cdn.aaai.org/KDD/1996/KDD96-037.pdf}.

\end{thebibliography}\endgroup

\end{document}